\documentclass[amssymb,nofootinbib]{revtex4-1}
\usepackage{amssymb}
\usepackage{amsmath}
\usepackage{graphicx}
\usepackage{subfig}
\usepackage{url}
\usepackage{dcolumn}
\usepackage{bm}

\DeclareMathOperator{\ev}{eV}

\DeclareMathOperator{\mev}{MeV}
\DeclareMathOperator{\gev}{GeV}

\newcommand{\be}{\begin{equation}}
\newcommand{\ee}{\end{equation}}
\usepackage{amssymb}
\usepackage[final]{pdfpages}
\usepackage{fullpage}
\usepackage{amsmath}
\usepackage{graphicx}
\usepackage{latexsym}
\usepackage{amsthm}
\usepackage{slashed}

\newcommand{\cM}{{\cal M}}

\newcommand{\cO}{{\cal O}}

\newcommand{\cH}{{\cal H}}

\newcommand{\Br}{{\rm Br}}

\newcommand{\half}{\frac{1}{2}}

\newcommand{\beq}{\begin{equation}}
\newcommand{\eeq}{\end{equation}}

\begin{document}

\begin{flushright}MCTP-12-11\\
\end{flushright}

\title{Neutrino Phenomenology in a 3+1+1 Framework}
\author{Eric Kuflik}
\affiliation{
Raymond and Beverly Sackler School of Physics and Astronomy, \\Tel-Aviv University, Tel-Aviv 69978, Israel
}
 \author{Samuel D. McDermott and Kathryn M. Zurek}
\affiliation{
Michigan Center for Theoretical Physics, Department of Physics, University of Michigan, Ann Arbor, MI 48109
}
\date{\today}

\begin{abstract}
\noindent
Evidence continues to grow in the MiniBooNE (MB) antineutrino mode supporting a low-energy excess compatible with the MB neutrino mode and possibly also confirming the results of the LSND experiment.  At least one sterile neutrino is required to explain the anomalies consistent with the observations of other experiments.   At the same time, there is a strong tension between the positive signals of LSND and MB and the null results of $\nu_e$ and $\nu_\mu$ disappearance experiments.  We explore a scenario, first proposed in \cite{Nelson:2010hz}, where the presence of an additional heavy sterile neutrino (with mass well above an eV) can alleviate tension between LSND, MB and the null results of disappearance experiments.  We compare and contrast this 3+1+1 scenario with the more standard 3+1 scenario and carry out global fits to all oscillation data including new 2011 MB $\bar{\nu}$ data.  We find that the tension can be somewhat alleviated and that a phenomenologically viable window for the heavy neutrino, consistent with rare decays and BBN constraints, can be found if the fifth neutrino has a mass of order $0.3 - 10 \gev$.  We also find, however, that the 2011 MB $\bar{\nu}$ data exacerbates the tension with null experiments in both the 3+1 and 3+1+1 models when the lowest energy bins are included, resulting in little improvement in the global fit.  We also discuss the implications of an additional neutrino for the reactor and gallium anomalies, and show that an oscillation explanation of the anomalies is disfavored by cosmological considerations, direct searches, and precision electroweak tests.
\end{abstract}

\maketitle

\tableofcontents

\section{Introduction}

Neutrino masses imply the presence of new states that can generate neutrino mass terms consistent with the standard model (SM) $SU(2)$ gauge symmetry.   Since the observed neutrino mass splittings are tiny, the standard way to implement the new states is to decouple them by giving them large masses.   At energies relevant for neutrino experiments, this gives rise to a new higher-dimension operator which generates neutrino masses, presumably at the scale of grand unification.  Since the dynamics of the new physics is decoupled, however, this mechanism for neutrino mass generation can never be directly tested.

Recent experimental hints have, on the other hand, suggested that there may be new dynamics in the neutrino sector at a much lower scale, leading to the possibility of probing the neutrino mass generation mechanism directly. The LSND \cite{hep-ex/0104049} and MiniBooNE (MB) \cite{AguilarArevalo:2007it,AguilarArevalo:2008rc,AguilarArevalo:2010wv} experiments both have reported results consistent with oscillations through a new sterile neutrino mass eigenstate with a splitting that is larger than the splittings that control the oscillations of the SM neutrinos.  The SM mass splittings are fixed by the observations in atmospheric and solar neutrino experiments to be $\cO(10^{-3} \ev^2)$ and $\cO(10^{-5} \ev^2)$, respectively.  By contrast, the results from  the LSND and MB $\bar{\nu}_\mu \to \bar{\nu}_e$ and $\nu_\mu \rightarrow \nu_e$ searches are consistent with a mass-squared splitting roughly between 0.1 and $1\ev^2$, which would require a new, heavier neutrino mass eigenstate.

The simplest extension of the SM that can satisfy these requirements is a single sterile neutrino (the 3+1 scheme). The existence of such a neutrino in the LSND and MB preferred mass region is, however, disfavored by global fits to the data \cite{arXiv:1103.4570}, since null searches for neutrino disappearance tightly constrain the mixing angles needed to produce the LSND and MB signals. Measurements of $\bar{\nu}_e$ fluxes from nuclear reactors and $\bar{\nu}_\mu$ fluxes from beam dump experiments can be combined to reject the relatively large mixing angles required by LSND and MB. As we show below, this statement remains true even using the new reactor flux predictions as inputs. Thus, a new neutrino capable of explaining the combined neutrino oscillation data enters a very constrained parameter space.

\begin{figure}[t]
\begin{center}
\includegraphics[width=4.5in]{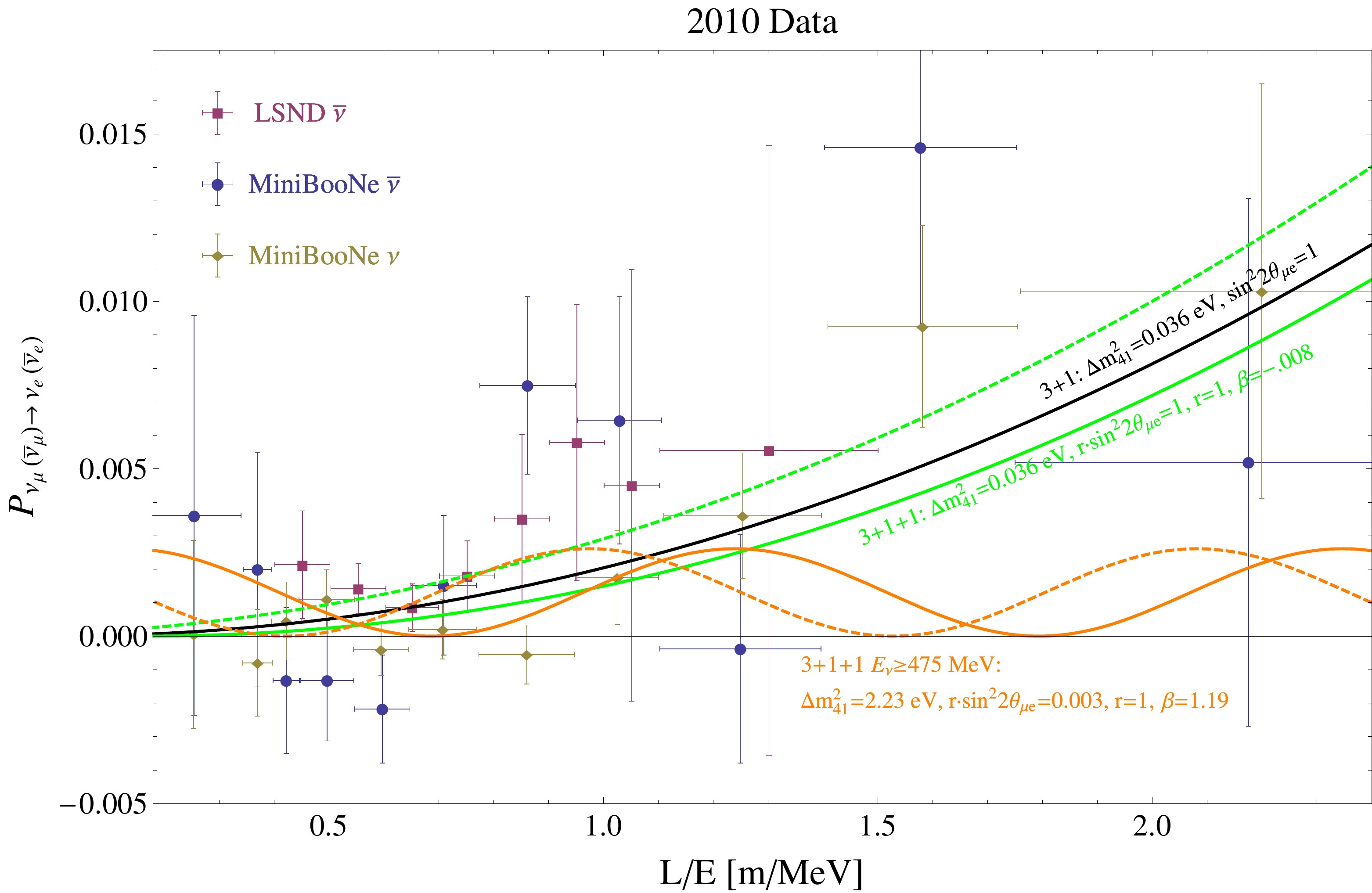} \\~\\
\includegraphics[width=4.5in]{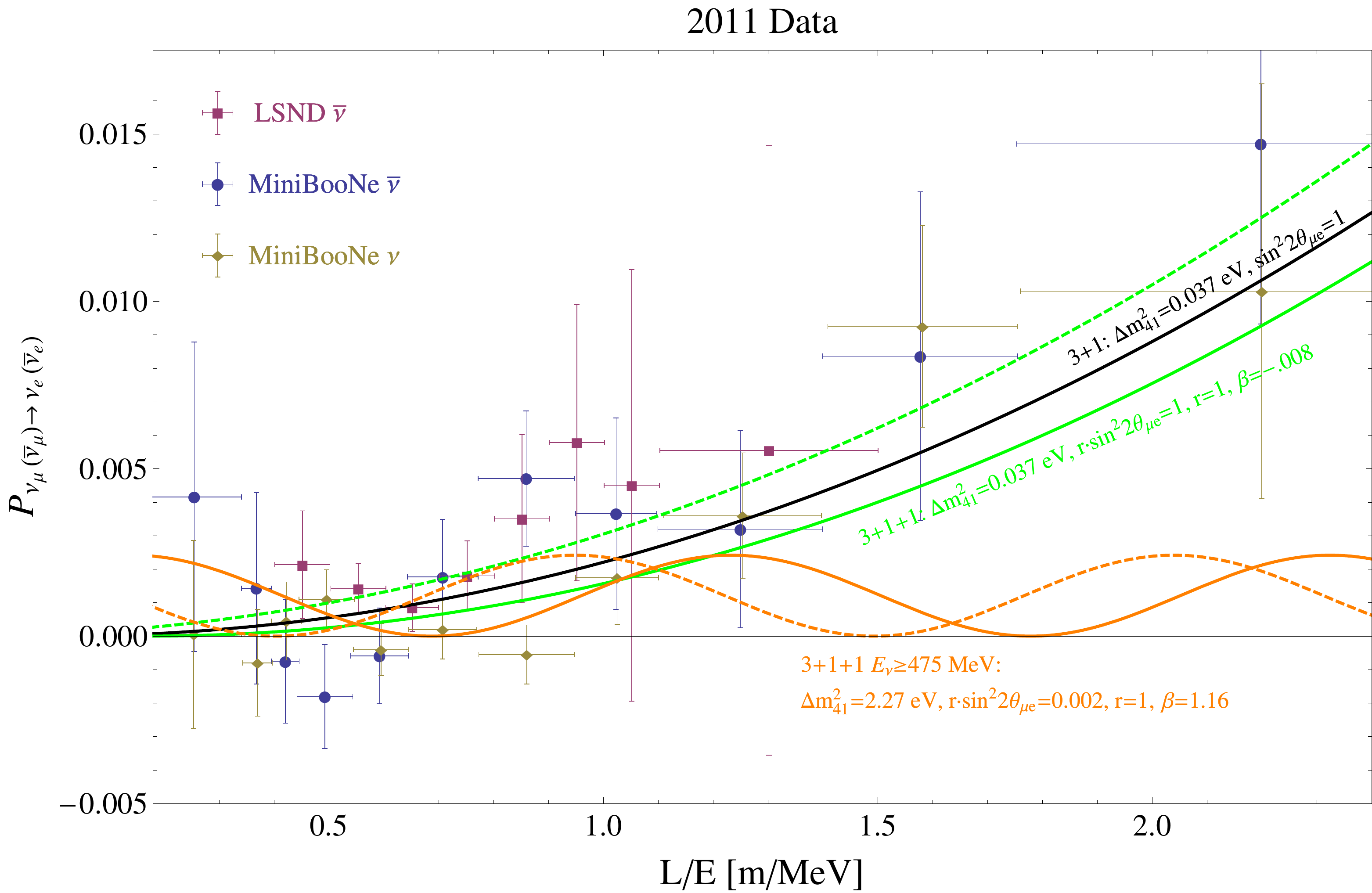}
\caption{Comparison of 2010 (upper panel) and 2011 (lower panel) MB $\bar{\nu}$ $L/E$ data with MB $\nu$ and LSND $\bar{\nu}$ $L/E$ data.  In both panels, the MB $\nu$ data is taken from \cite{AguilarArevalo:2008rc}.  In the upper panel, the MB $\bar{\nu}$ data is taken from \cite{AguilarArevalo:2010wv}, while in the lower panel the MB $\bar{\nu}$ data has been updated with the results of \cite{Huelsnitz}.  We show the best fit lines in the 3+1 scenario (black), the 3+1+1 scenario (green) using all data points, and the 3+1+1 scenario (orange) dropping the three low-energy data points so that the data is in the range $E_\nu>475\mev$. In all plots, $\nu$ lines are dotted and $\bar{\nu}$ lines are solid.
}
\label{oldvnew}
\end{center}
\end{figure}

In addition to these considerations, there are more complications facing the 3+1 hypothesis.  Early results from MB \cite{AguilarArevalo:2007it,AguilarArevalo:2010wv} suggested that such a 3+1 scheme might not have been compatible with the MB data alone, since the parameters needed to fit $\bar{\nu}_\mu \to \bar{\nu}_e$ and $\nu_\mu \rightarrow \nu_e$ appeared to be different: at high energy the MB anti-neutrino mode favored oscillations and was in better agreement with the $\bar{\nu}_\mu \to \bar{\nu}_e$ LSND data, while $\nu_\mu \rightarrow \nu_e$ data was consistent with a null result.\footnote{A low-energy excess in the $\nu_\mu$ channel was initially suspected of being a systematic effect \cite{AguilarArevalo:2007it} and was reported to be incompatible with a neutrino oscillation interpretation \cite{AguilarArevalo:2008rc}.} The addition of a second sterile neutrino (the 3+2 scheme) allows for CP violation which can reconcile differences in $\bar{\nu}_\mu \to \bar{\nu}_e$ and $\nu_\mu \rightarrow \nu_e$.  However, this scheme suffers from a similar tension between the null data and the positive signals and does little to ameliorate the difficulties of the 3+1 scheme \cite{arXiv:1103.4570,Maltoni:2007zf}.  In addition, new $\bar{\nu}$ data \cite{Huelsnitz} from MB, shown in Fig.~\ref{oldvnew}, indicates that the apparent difference between the $\nu$ and $\bar{\nu}$ modes is disappearing, thereby obviating one of the primary appeals of the 3+2 framework.

Still, the tension between the null results and the LSND and MB data persists.  In this paper we consider a simple scheme, proposed in~\cite{Nelson:2010hz}, designed to alleviate the tension between the LSND and MB positive signals and the null results from reactor and short baseline experiments.  This scenario requires a single light sterile neutrino with mass splitting in the MB and LSND range between 0.1 and 1 eV$^2$ and a second much heavier ($\Delta m^2 \gg 1000\ev^2$) neutrino whose oscillations are averaged over.  The heavier neutrino participates directly or indirectly in both disappearance and appearance experiments. Because most disappearance experiments have their first detector relatively far from the neutrino production point the heavy neutrino has undergone many oscillations before reaching the detector, and the effect of the heavy neutrino is to change the flux of the initial flavor neutrinos. If this flux is not precisely known, as is true in many reactor experiments, the experiment is relatively insensitive to oscillations through the heavy neutrino. Appearance experiments, by contrast, look for the appearance of a new flavor in a pure initial flavor beam, so they are sensitive to oscillations through the heavy neutrinos.  In this way, if the initial neutrino flux is not very well known in the disappearance experiments, new parameter space may open for the appearance experiments, giving rise to the possibility that the positive signals from LSND and MB are no longer in conflict with the results from otherwise null experiments.

The purpose of this paper is two-fold.  First, we examine the 3+1 scenario in light of the new MB $\bar{\nu}$ data. This improves the compatibility of the combined appearance data within the 3+1 framework, but we find that the best fit region shifts considerably to larger mixings and smaller mass splittings, which increases the tension with the null experiments. Second, we explore the phenomenology of, and present constraints on, the ``3+1+1" framework of~\cite{Nelson:2010hz,Fan:2012ca}.  We will examine exactly how and to what extent the fifth neutrino is able to have an effect on the allowed parameter space of the fourth neutrino.

The outline of this paper is as follows.  We begin by establishing our notation and conventions.  We then carry out fits for the  3+1 scenario in light of new data from MB, complete with constraints from a diverse set of null experiments.  We then turn to discussing the parameter space for the 3+1+1 scenario with respect to neutrino experiments, before analyzing in detail the constraints from BBN, astrophysics and rare decays, which constrain the fifth neutrino to have a mass $\sim 0.3 -10 \gev$. In section IV we present aspects of some models that explicitly realize the features of the 3+1+1 scenario, and we conclude in section V.

\section{Phenomenology Of Sterile Neutrino Models}\label{sec:31}

We establish our notation and contrast the 3+1 framework (see {\it e.g.} \cite{Maltoni:2007zf,arXiv:1103.4570,Giunti:2011cp,Bhattacharya:2011ee}) with the 3+1+1 scheme \cite{Nelson:2010hz,Fan:2012ca}. We will lay out some conventions for discussing these models, leaving a more complete discussion of statistical methods and derivation of the oscillation formulae in the 3+1+1 scenario to the appendix.

We aim to examine the oscillation appearance and disappearance data in depth, with a specific emphasis on the new conclusions to be drawn from some recently presented preliminary MB data~\cite{Huelsnitz}. This new data is in better agreement with the LSND data and prefers a sterile neutrino with lower mass and more substantial mixing than indicated by the earlier MB data.

\subsection{Conventions}

We parameterize the mass mixing by
\beq
\nu_{\alpha} = \sum_{i=1}^{N}  U_{ \alpha i} n_i,
\eeq
where $\nu_{\alpha}$ are the neutrino flavor eigenstates, which include the 3 left-handed (active) neutrinos of the SM plus any $SU(2)$-singlet (sterile) neutrinos; $U_{ \alpha i}$ are the elements of a unitary $N \times N$ matrix that diagonalizes the neutrino mass matrix and causes mixing between the neutrino flavor eigenstates; and $n_i$ are the neutrino mass eigenstates with mass $m_i$ ordered by increasing mass.

Since we are focusing on a 3+1 scheme (with a single light neutrino) and a 3+1+1 scheme (with one light and one heavy neutrino) we are generally interested in oscillation probabilities where all but one of the mass eigenstates are easily kinematically accessible. From Eq.~\eqref{oscform}, the probability of detecting $\nu_\beta$ in a $\nu_\alpha$ beam is
\beq \begin{array}{rcl}
P_{\nu_\alpha \to \nu_\beta}&=& \delta_{\alpha \beta} \left[ 1 +2(a -1) |U_{\alpha 5}| |U_{ \beta 5}| \right] + (1-a) |U_{\alpha 5}|^2 |U_{\beta 5}|^2     \\
&&  ~- 4  \sum_{5>i>j} \Re \{ U_{\alpha i}^* U_{\beta i}  U_{\alpha j} U_{\beta j}^* \} \sin^2 x_{ij}- 4 a  \sum_{j=1}^4 \Re \{ U_{\alpha 5}^* U_{\beta 5}  U_{\alpha j} U_{\beta j}^* \}  \sin^2 x_{5j}   \\
&&~ - 2 \sum_{5>i>j} \Im \{ U_{\alpha i}^* U_{\beta i}  U_{\alpha j} U_{\beta j}^* \}\sin 2 x_{ij}  - 2 a \sum_{j=1}^4 \Im \{ U_{\alpha 5}^* U_{\beta 5}  U_{\alpha j} U_{\beta j}^* \}\sin 2 x_{5j} .
\end{array} \eeq
Here $x_{ij}=\Delta m_{ij}^2 L/4E=1.27 \frac{(m_i^2-m_j^2) L/E}{\ev^2 {\rm m}/ \mev}$, where $L$ is the distance the neutrino has traveled and $E$ is the neutrino energy. Since $n_5$ will be much heavier than the other neutrinos, accounting for the possibly suppressed production of and oscillation through $n_5$ requires a phase space factor $a$ that interpolates from 0 (kinematically forbidden) to 1 (phase space fully accessible) as a function of the neutrino energy. For the short baselines and high energies of the experiments under consideration it will be a good approximation to take $x_{ij} \simeq 0$ for $i$ and $j=1,2,3$, and this formula simplifies considerably.
For instance, the probability for disappearance of flavor $\alpha$ is
\beq \label{disappsimple}
1-P_{\nu_\alpha \to \nu_\alpha} = \sin^2 2 \theta_{\alpha 4}  \sin^2 x_{41} + 2 |U_{\alpha 5}|^2 \left(1  - \frac{a+1}{2} |U_{\alpha 5}|^2 \right) ,
 \eeq
where we define $\sin^2 2 \theta_{\alpha 4}=4 |U_{\alpha 4}|^2 ( 1- |U_{\alpha 4}|^2  -|U_{\alpha 5}|^2 )$ and we assume that the characteristic oscillation length associated with $\Delta m_{51}^2$ is so short that $\sin^2 x_{51} \to \half$ holds over the volume of the detector. Experiments that probe disappearance of $\nu_e$ are carried out at reactors and in solar neutrino searches, while $\nu_\mu$ disappearance is probed by beam dump and atmospheric neutrino experiments.

Following \cite{Nelson:2010hz}, the probability for $\nu_e$ appearance in a $\nu_\mu$ beam, measured by LSND and MB among others, simplifies. From Eq.~\eqref{genapp}
\be \label{appmaster}
P_{\nu_\mu (\bar{\nu}_\mu) \to \nu_e (\bar{\nu}_e)} =  \sin^2  2 \theta_{\mu e} \sin^2 \left(x_{41} \pm \beta \right) + \kappa,
\ee
with the definitions
\be
\begin{array}{rcl}
 \sin^2 2 \theta_{\mu  e} &=&   4 \left|U_{\mu 4}\right|^2 \left|U_{e4}\right|^2 r   \\
 \kappa &=&  \left|U_{\mu 4}\right|^2 \left|U_{e4}\right|^2 \left\{ (1-r)^2 + a \left[(1-r)^2 + 4 r \sin^2\beta \right] \right\} \\
\end{array} \label{appdefs}
\ee
 where $+$($-$) is for $\nu$ ($\bar{\nu}$) oscillations,
\be
\begin{array}{rcl} \label{rdef}
 r &\equiv&  \left|U_{\mu 4}^* U_{e4} + U_{\mu 5}^* U_{e5}\right| / \left|U_{\mu 4}^* U_{e4} \right|    \\
 \beta &\equiv& \frac{1}{2} \tan^{-1}\left(\frac{\sin\phi |U_{e 5}||U_{\mu 5}| }{  |U_{e 4}| |U_{\mu 4}|+\cos\phi |U_{e 5}||U_{\mu 5}| }\right)
\end{array}
\ee
and $\phi\equiv \arg\left( \frac{U_{e5}U_{\mu 5}^*}{U_{e 4}U_{\mu 4}^*}\right)$. $\beta$ is the CP-odd parameter that can account for differences in $\nu$ and $\bar{\nu}$ oscillations. The 3+1 model can be recovered in the limit $U_{e5}$ and $U_{\mu5} \to 0$, or $r=1$ and $\kappa=\beta=0$. We emphasize that the sensitivity to the mixings with $n_5$ is such that even the limit $a \to 0$ produces nontrivial oscillation effects.

The 3+1+1 model is capable of opening parameter space closed by 3+1 models because of the possibility of CP violation and because in general we can have $r>1$. In the small mixing, CP-conserving limit the effect of $r$ is multiplicative because we may make the approximation $\sin^2 2\theta_{\mu e} \simeq r \sin^2 2\theta_{e4} \sin^2 2 \theta_{\mu 4}/4$, and the limits from disappearance experiments can be made compatible with larger appearance mixings if one has $r>1$. However, we will show that because of the presence of the term that depends on $|U_{e5}|^2$ in Eq.~\eqref{disappsimple} the constraints on the mixings with $n_5$ are almost as strong as the constraints on the mixings with $n_4$. This forces $r$ to be close to 1 for most of the interesting parameter space, and $r$ is not effective in practice for reconciling the appearance and disappearance experiments.

\subsection{Fits to Neutrino Appearance Anomalies}

\begin{figure}[t]
\begin{center}
\includegraphics[width=.45\textwidth]{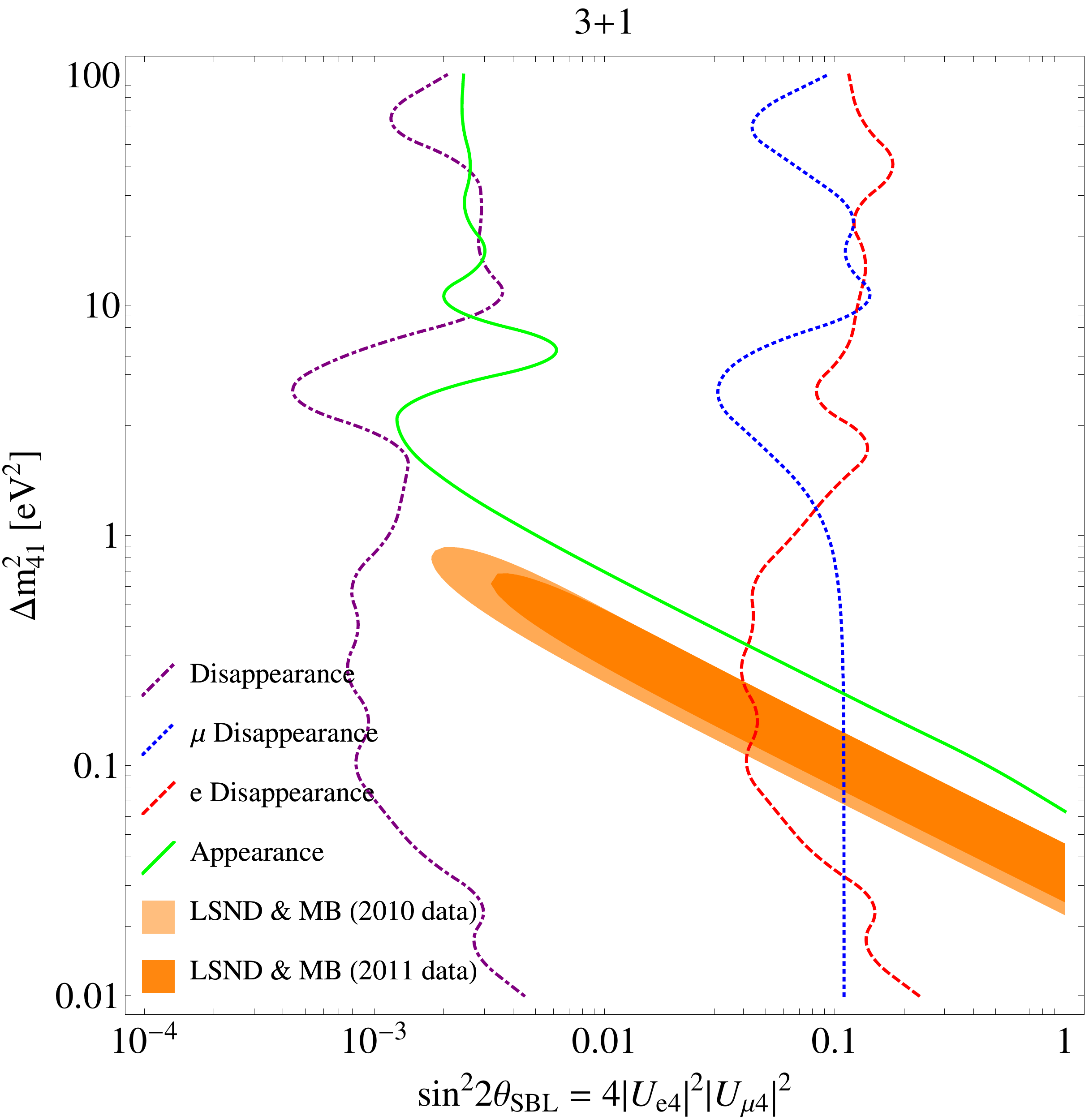}~~~
\includegraphics[width=.45\textwidth]{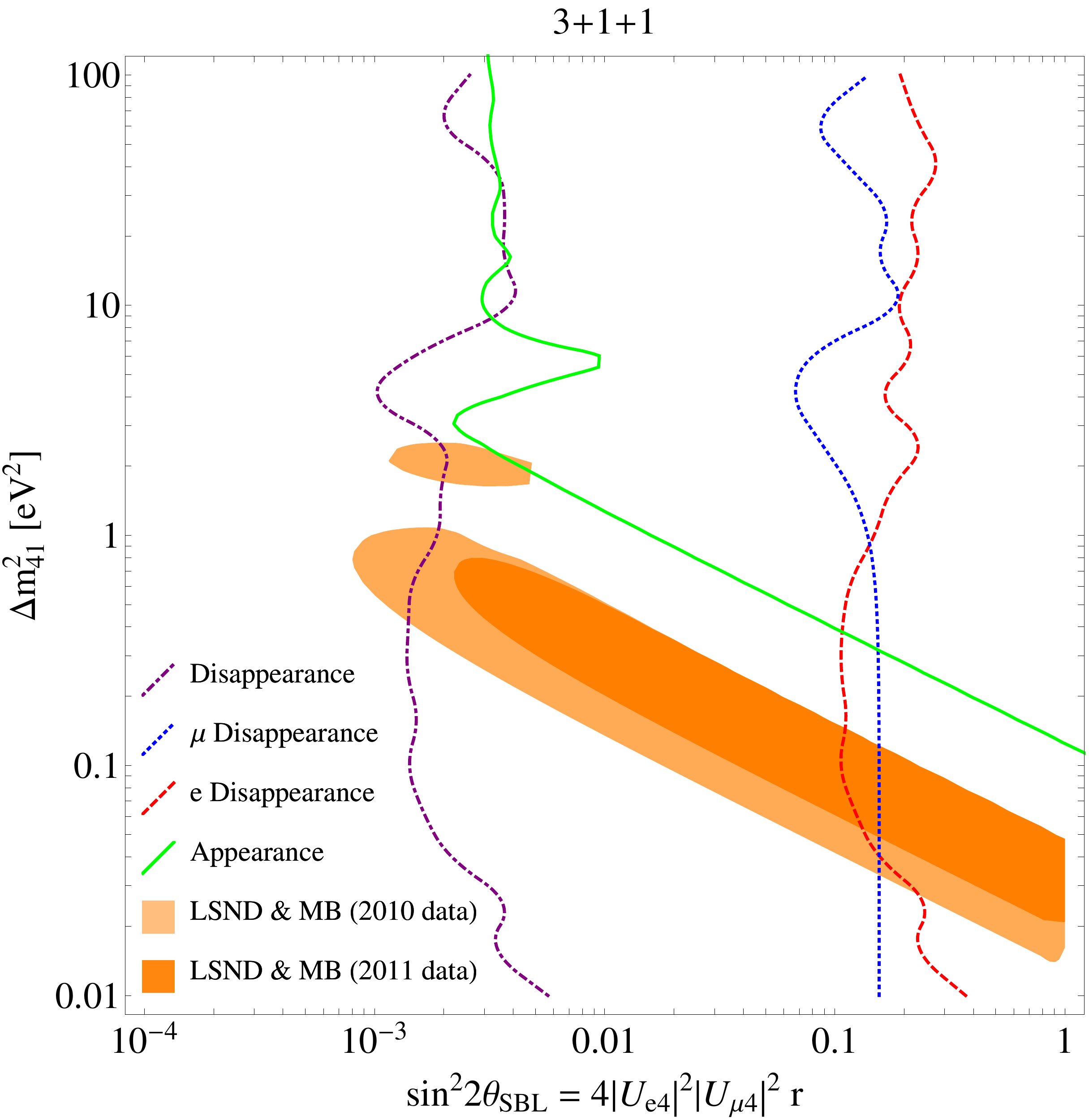}
\caption{Fits in the 3+1 (left) and 3+1+1 (right) neutrino models. We also contrast the allowed regions using the 2010 (light-orange) and the 2011 (dark-orange) MB $\bar{\nu}$ data. In both panels we show the appearance allowed region at 99\% as well as the appearance null result and disappearance null result  exclusion curves at 99\%.  There is significant tension with the disappearance experiments and oscillations reported by LSND and MB for both the 3+1 and 3+1+1 scenarios with the 2011 data.}
\label{AA}
\end{center}
\end{figure}

Fits to the 3+1 and 3+1+1 frameworks with all relevant data are shown in Fig.~\ref{AA}; we display them side by side to enhance comparisons of the fits.  In each panel we superimpose the results using the 2010 \cite{AguilarArevalo:2010wv} and 2011 \cite{Huelsnitz} $\bar{\nu}$ data from MB. We use the 2009 data \cite{AguilarArevalo:2008rc} for the $\nu$ mode for all fits. The best fit to the data, using either the 2010 or 2011 MB $\bar{\nu}$ data, indicates a new sterile neutrino described by a mass splitting $\Delta m_{41}^2 \sim \cO(0.03 \ev^2)$ and a mixing angle roughly of size $1$, although the $\chi^2$ is relatively shallow and is consistent with mass splitting $\Delta m_{41}^2 \sim \cO(0.5 \ev^2)$ and mixing angle $\sim \cO(3\times10^{-3})$.  These values differ from those we would find if we omitted the low-energy MB $\nu$ and $\bar{\nu}$ points.  As we discuss in more detail below, dropping these points reduces the significance of the signal so that the data are compatible with no oscillations at the 99\% level, as noted in \cite{Huelsnitz}.  Because much of the significance of the fit to oscillations is derived from events with $E_\nu < 475\mev$, we do not omit these points in our fits.

In principle, both appearance and disappearance oscillation experiments can bound the LSND and MB preferred region. We consider null appearance searches at KARMEN \cite{Armbruster:2002mp}, E776 \cite{Borodovsky:1992pn},  NOMAD \cite{Astier:2003gs}, CCFR \cite{Romosan:1996nh}, and NuTeV \cite{Avvakumov:2002jj},  and we find that the preferred region using the new MB $\bar{\nu}$ data is no longer in tension with these searches, due to the lower-mass preferred region. Although the LSND and MB oscillation results are not strongly constrained by the null appearance searches, the mixing angle probed by the appearance experiments can be tightly constrained by combining the results of $\nu_e$ and $\nu_\mu$ disappearance experiments. The disappearance experiments independently constrain $\sin^2  2 \theta_{e 4}$ and  $\sin^2  2 \theta_{\mu 4}$, and in a 3+1 scenario with small mixing angles we can approximate $\sin^2 2 \theta_{\mu e} \simeq \sin^2 2 \theta_{e4} \sin^2 2 \theta_{\mu4}/4$, so we obtain limits on the LSND and MB parameter space by combining the two sets of constraints. Details of how these constraints are combined are given in the appendix. The $\nu_e$ disappearance constraints include short-baseline reactor experiments with new reactor flux predictions \cite{Mention:2011rk}\footnote{The new reactor flux has been reported to reflect oscillations of a sterile neutrino, but we find that it is not consistent with our preferred region, and we use the reactor data as a constraint. We discuss a possible resolution to this anomaly below.} as well as constraints from the ratio of flux observed in the Bugey 40 m and 15 m detectors \cite{Declais:1994su}. Disappearance of $\nu_\mu$ is constrained by CDHS \cite{Dydak:1983zq} and CCFR \cite{Stockdale:1984ce} at high mass. We also take into account mass-independent unitarity constraints arising from the maximal measurement of the atmospheric ($\nu_\mu$) \cite{Wendell:2010md} disappearance mixing angles made by the Super-Kamiokande experiment. In the appendix we show that this leads to $|U_{\mu 4}|^2+|U_{\mu 5}|^2 < 0.0175 ~(0.0274)$ at 90\% (99\%) confidence.

The best fit oscillation statistics for the 3+1 scheme are given in Table~\ref{PGtab1}, where we show the $\chi^2_{\rm min}$ values for the disappearance and appearance data sets individually as well as the $\chi^2_{\rm min}$ for the global data set. The value of the $\chi^2_{\rm min}/$DOF for the global fit does not indicate a bad fit to the data (as noted in, {\it e.g.}, \cite{Giunti:2011cp}) but the $\chi^2_{\rm PG}$ \cite{Maltoni:2003cu} value for the different data sets is very high, which indicates that the data as a whole are not compatible. This is reflected in Fig.~\ref{AA}, which shows the ``disappearance" curve ruling out the ``LSND \& MB" allowed region.
\begin{table}[b]
\begin{tabular}{cc}
2010 Data & 2011 Data \\
ÊÊÊÊ\begin{tabular}{lrr}
ÊÊÊÊÊÊÊÊÊÊÊÊÊÊÊÊÊÊÊÊÊÊ & $\chi^2_{\rm min}$ & bins  \\ \hline\hline
ÊÊÊÊÊÊÊÊDisappearance		& 25.4		& 49\\
ÊÊÊÊÊÊÊÊAppearance		& 0.20		& 5  \\
ÊÊÊÊÊÊÊÊLSND + MB		& 32.1		& 30\\
ÊÊÊÊÊÊÊÊEverythingÊÊÊÊÊÊÊÊÊÊÊÊÊÊ 	& 75.4		& 84
ÊÊÊÊ\end{tabular} & Ê\begin{tabular}{lrr}
ÊÊÊÊÊÊÊÊÊÊÊÊÊÊÊÊÊÊÊÊÊÊÊÊ & $\chi^2_{\rm min}$ & bins \\ \hline\hline
ÊÊÊÊÊÊÊÊDisappearance		& 25.4		& 49\\
ÊÊÊÊÊÊÊÊAppearance		& 0.20		& 5  \\
ÊÊÊÊÊÊÊÊLSND + MB		& 24.0		& 30\\
ÊÊÊÊÊÊÊÊEverythingÊÊÊÊÊÊÊÊÊÊÊÊÊÊ 	& 72.9		& 84
ÊÊÊÊ\end{tabular}
\\~&~\\
ÊÊÊÊÊÊ\begin{tabular}{c}
ÊÊÊÊ$ \chi^2_{\rm PG} = \left( \sum \chi^2 \right)_{\rm min} - \sum \chi^2_{\rm min} = 17.7$ \\
ÊÊÊÊÊÊÊÊp-value  $=  5.02  \times 10^{-4}~(3.48~\sigma)$ÊÊÊÊÊÊÊÊÊÊÊÊÊÊÊÊÊÊ \\
ÊÊÊÊ\end{tabular} & \begin{tabular}{c}
ÊÊÊÊÊÊÊÊ$\chi^2_{\rm PG} = \left( \sum \chi^2 \right)_{\rm min} - \sum \chi^2_{\rm min}  =23.3$\\
ÊÊÊÊÊÊÊÊp-value $=3.44 \times 10^{-5}~ (4.14~\sigma)$ÊÊÊÊÊÊÊÊÊÊÊÊÊÊÊÊÊÊ \\
ÊÊÊÊ\end{tabular}
\end{tabular}
\caption{ Fits to the 3+1 framework using 2010 and 2011 $\bar{\nu}$ data. With the new MB data, the appearance and disappearance experiments disagree at more than the $4 \sigma$ level.} \label{PGtab1}
\end{table}
The new data play an important role in shifting the preferred region and increasing the tension between appearance and disappearance: we find that the appearance data on its own is marginally more self-consistent when incorporating the 2011 MB $\bar{\nu}$ data ($\chi^2_{\rm min,LSND+MB2011} = 24.0$) instead of the 2010 data ($\chi^2_{\rm min,LSND+MB2010} = 32.1$), while the parameter goodness of fit becomes slightly worse ($\chi^2_{\rm PG,2010}=17.7$ and $\chi^2_{\rm PG,2011}=23.3$).  This is a result of a more significant departure from the null oscillation hypothesis at large $L/E$ in the 2011 data, which is compatible with the excess at low energy found in the MB $\nu$ data, as can be seen in Fig.~\ref{oldvnew}. Due to the increased power at large $L/E$, our global appearance region is at somewhat lower mass and higher mixing than shown by previous global fits ({\it e.g.}, \cite{arXiv:1103.4570}). We conclude that the tension between the positive signals and the null searches indicates that a single sterile neutrino is very unlikely to explain the entirety of the collected data.

In the right panel of Fig.~\ref{AA} we show the results of a similar analysis performed in the 3+1+1 framework, where the fit region differs from the the 3+1 case because of the CP-odd phase $\beta$ and the multiplicative factor $r$.
\begin{figure}[t]
\begin{center}
\includegraphics[width=1\textwidth]{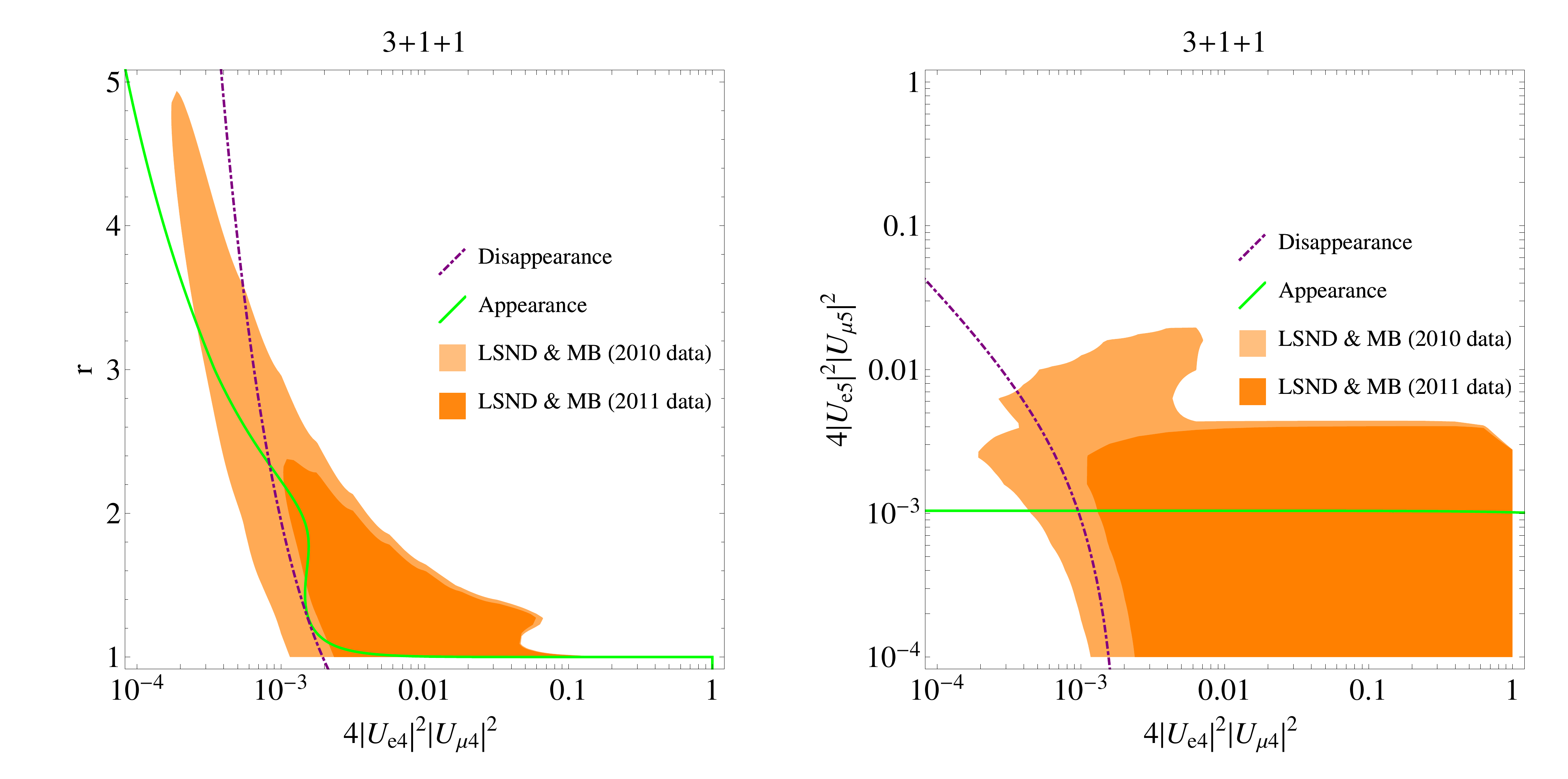}
\caption{Constraints on $r$ as a function of $|U_{e4}|^2 |U_{\mu 4}|^2$ for $0.04\ev^2<\Delta m_{41}^2<6\ev^2$. We see that $r$ is close to 1 in the appearance preferred region, and has limited ability to reduce the tension with null experiments.}
\label{rfig}
\end{center}
\end{figure}
The parameter $r$ represents a potentially significant handle in the 3+1+1 framework, since it can give a multiplicative enhancement of the appearance angle compared to the disappearance angles. For small $\beta$, $r$ effectively measures the magnitude of the mixings with $n_5$, and to obtain the desired enhancement over the 3+1 scheme we need the mixings with $n_5$, and thus the value of $r$, to be greater than 1. However, as shown in the left panel of Fig.~\ref{rfig}, we find that $r$ is bounded by the null experiments to be very close to 1 for most of the values of $|U_{e4}|^2$ and $|U_{\mu 4}|^2$ favored by LSND and MB. This is a consequence of the ``zero-distance" effect \cite{Langacker:1988ur}, which allows for the oscillation of neutrino flavors at arbitrarily low distances. The zero-distance effect can manifest itself in two ways in the experiments in consideration. First, disappearance experiments bound the {\it sum} of the mixing angles due to this effect, as in Eq.~\eqref{disappsimple}, which forces either $|U_{\alpha 5}|^2$ or $|U_{\alpha 4}|^2$  to be small. Thus, disappearance experiments constrain $r$ to be very close to 1 for large $|U_{\alpha 4}|^2$, as is true in the appearance preferred region. This effect is shown in the left panel of Fig.~\ref{rfig}, where $r$ is seen to be essentially compatible with 1 in the entire appearance preferred region.  The second way the zero-distance effect would be visible is as a positive offset in the appearance probability. However, the appearance experiments exhibit transition probabilities roughly of order 0.5\%, which allows them to place their own firm upper bound on $|U_{e5}|^2|U_{\mu 5}|^2$, as is visible in the right panel of Fig.~\ref{rfig}.  In other words, we find that $r$ is negligibly effective in reconciling the appearance and disappearance data sets.

\begin{figure}[t]
\begin{center}
\includegraphics[width=1\textwidth]{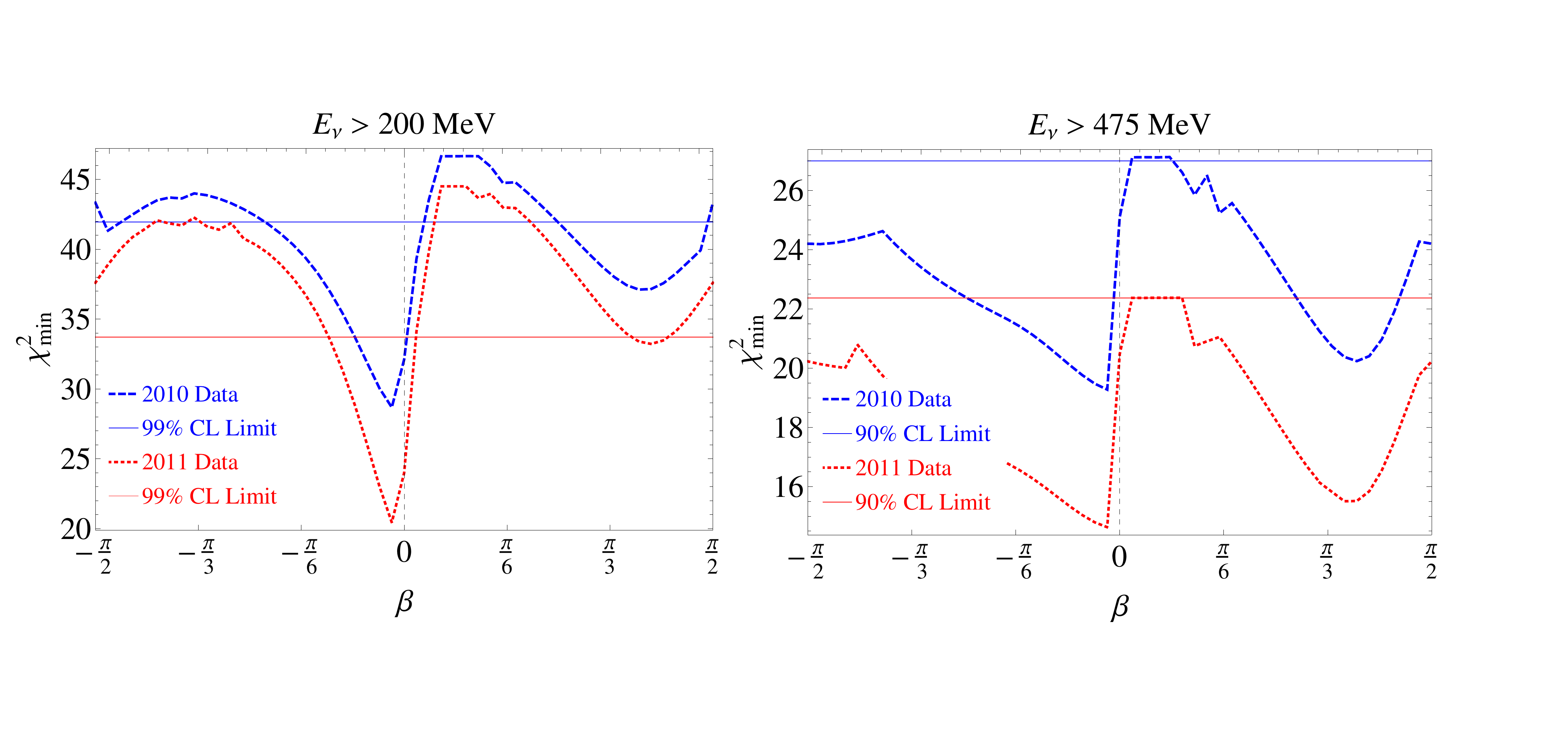}
\caption{$\chi^2$ as a function of the CP-odd parameter $\beta$, with (left) and without (right) the MB $\nu$ and $\bar{\nu}$ data for $200\mev<E_\nu<475\mev$. We show fits utilizing both the 2010 and 2011 MB $\bar{\nu}$ data, and we show the 90\% or 99\% allowed value from the $\Delta \chi^2$ test that we use.}
\label{beta}
\end{center}
\end{figure}

The other potential advantage of the 3+1+1 framework is the possibility of CP violation, but we also find that this is not very effective in reducing the tension with the null experiments. As shown in the left panel of Fig.~\ref{beta}, the $\chi^2$ has a pronounced preference for a small nonzero value of $\beta$, and the sharpness of this feature means that the extra parameter freedom is largely unimportant in defining our preferred region. When we drop the MB $\nu$ and $\bar{\nu}$ data points below $475\mev$, as advocated in the initial MB data release \cite{AguilarArevalo:2007it}, we find that CP violation has a much more significant impact on the fit.  This is because the $\chi^2$ is substantially flatter as a function of $\beta$ and exhibits two rather broad and nearly degenerate minima, as shown in the right panel of Fig.~\ref{beta}. This in turn is a result of the mostly flat spectrum of the $\nu$ and $\bar{\nu}$ data points in the region $E_\nu > 475 \mev$, which can be made compatible at the 1 $\sigma$ level for a wide range of low-mixing-angle oscillations when CP violation is allowed. The ultimate effect of the shallower $\chi^2$ is to reduce the preference for a particular mass or mixing, which opens up a wider range of parameter space and broadens the preferred region.  When we perform fits analogous to those in Fig.~\ref{AA} for the data with $E_\nu > 475 \mev$, we find that the significance of the signal drops so drastically that the remaining data are consistent with no oscillations at the 99\% level, as noted in \cite{Huelsnitz}.  We show the CP-violating best fits to the $E_\nu>475\mev$ data alongside the best fits to the $E_\nu>200 \mev$ data in Fig.~\ref{oldvnew}.

\begin{table}[h]
\begin{tabular}{cc}
2010 Data & 2011 Data \\
ÊÊÊÊ\begin{tabular}{lrr}
ÊÊÊÊÊÊÊÊÊÊÊÊÊÊÊÊÊÊÊÊÊÊ & $\chi^2_{\rm min}$ & bins  \\ \hline\hline
ÊÊÊÊÊÊÊÊDisappearance		& 24.6		& 49\\
ÊÊÊÊÊÊÊÊAppearance		& 0.20		& 5  \\
ÊÊÊÊÊÊÊÊLSND + MB		& 28.3		& 30\\
ÊÊÊÊÊÊÊÊEverythingÊÊÊÊÊÊÊÊÊÊÊÊÊÊ 	& 74.4		& 84
ÊÊÊÊ\end{tabular} & Ê\begin{tabular}{lrr}
ÊÊÊÊÊÊÊÊÊÊÊÊÊÊÊÊÊÊÊÊÊÊÊÊ & $\chi^2_{\rm min}$ & bins \\ \hline\hline
ÊÊÊÊÊÊÊÊDisappearance		& 24.6		& 49\\
ÊÊÊÊÊÊÊÊAppearance		& 0.20		& 5  \\
ÊÊÊÊÊÊÊÊLSND + MB		& 19.4		& 30\\
ÊÊÊÊÊÊÊÊEverythingÊÊÊÊÊÊÊÊÊÊÊÊÊÊ 	& 73.2		& 84
ÊÊÊÊ\end{tabular}
\\~&~\\
ÊÊÊÊÊÊ\begin{tabular}{c}
ÊÊÊÊ$ \chi^2_{\rm PG} =\left( \sum \chi^2 \right)_{\rm min} - \sum \chi^2_{\rm min} = 21.3$ \\
ÊÊÊÊÊÊÊÊp-value  $=  1.6  \times 10^{-3}~(3.16~\sigma)$ÊÊÊÊÊÊÊÊÊÊÊÊÊÊÊÊÊÊ \\
ÊÊÊÊ\end{tabular} & \begin{tabular}{c}
ÊÊÊÊ$ \chi^2_{\rm PG} =\left( \sum \chi^2 \right)_{\rm min} - \sum \chi^2_{\rm min}  =29.0$\\
ÊÊÊÊÊÊÊÊp-value $=6.0\times 10^{-5}~ (4.00~\sigma)$ÊÊÊÊÊÊÊÊÊÊÊÊÊÊÊÊÊÊ \\
ÊÊÊÊ\end{tabular}
\end{tabular}
\caption{ Results of fits to the 3+1+1 framework using 2010 and 2011 MB $\bar{\nu}$ data. With the new data, the appearance and disappearance data sets still disagree at about the $4 \sigma$ level, with only slight improvement over the 3+1 case.} \label{PGtab2}
\end{table}

Finally, we give the best fit statistics for the 3+1+1 model in Table~\ref{PGtab2}, taking $E_\nu>200\mev$ as usual. As in the 3+1 case, the 2011 MB $\bar{\nu}$ data provides slightly more agreement in the combined appearance data than the 2010 data. Again, the global fit to the data gives an acceptable $\chi^2/$DOF, but the PG test underscores the point that the data sets are incompatible. The p-value for the $\chi^2_{\rm PG}$ in the 3+1+1 case is slightly lower than in the 3+1 model for both the 2010 and 2011 MB $\bar{\nu}$ data. This suggests some improvement in agreement, but with the new data the tension remains at the $4\sigma$ level in both the 3+1 and 3+1+1 cases.

\section{Phenomenology of the Heavy Neutrino $n_5$ in the 3+1+1 Scheme}\label{sec:nu5bds}

We discuss the parameter space of interest for the heavy neutrino, $n_5$, that appears in the 3+1+1 framework. If $n_5$ is a Majorana neutrino the neutrinoless double beta decay constraints are extremely restrictive \cite{Atre:2009rg}, so we will take $n_5$ to be a Dirac state.

We begin by showing how some experimental anomalies recently reported at low significance could plausibly be explained by the existence of this heavy mass eigenstate. Then we proceed to place constraints on its parameter space. Tension with BBN constraints  forces us to the regime where $m_5 \gtrsim 1\mev$. Above this region constraints from SN1987A, pion and kaon decays, beam dump experiments, and nonobservation of $\mu \to e \gamma$ enter. We find that if one wishes to use $n_5$ to explain experimental anomalies, only a small window with $m_5 \sim \cO(1\gev)$ and mixing $|U_{e5}||U_{\mu 5}| \sim \cO(10^{-2})$ is allowed, with the additional requirement that $n_5$ be stable on collider timescales or decay to non-SM final states.

\subsection{The Gallium and Reactor Anomalies}\label{sec:RGC}

In recent years, several experiments have reported observing anomalously low neutrino fluxes.  Since these anomalies include many different energy and distance scales and exhibit no $L/E$ dependence, we do not include them in the fits to $n_4$, whose mass splitting will cause visible oscillations at these experiments. Instead, we fit to this data with the heavy neutrino $n_5$ whose oscillations are averaged over in all experiments. A somewhat more detailed discussion of the anomalies we fit with $n_5$ oscillations is given in the appendix.

The most statistically significant of these position- and energy-independent anomalies is the reactor antineutrino anomaly \cite{Mention:2011rk,Huber:2011wv} (RAA), where the global average of the observed $\bar{\nu}_e$ flux is less than anticipated by a factor $R_{\rm RAA}=0.943 \pm 0.023.$ In addition, anomalously low measurements of the $\nu_e$ scattering gallium cross-section \cite{gallium:ANOMALY}, the gallium anomaly (GA), may indicate disappearance of $\nu_e$. The deficit based on four measurements of the average $\nu_e$ scattering cross section from the process $\nu_e + {}^{71} {\rm Ga} \to {}^{71}{\rm Ge} + e^-$ is, with correlated errors taken to be those in  \cite{Mention:2011rk}, $R_{\rm GA}=0.86 \pm 0.06$.  Independent measurements of the strength of the relevant Gamow-Teller transitions \cite{Frekers:2011zz} supports the conclusion that this deficit might be due to averaged oscillations of a heavy neutrino. Finally, measurements of the energy dependence of the scattering cross section in the process $\nu_e + {}^{12}{\rm C \to {}^{12} N_{g.s.}} + e^-$  \cite{carbon:KARMENLSND} are very mildly discrepant with the cross-section predictions \cite{Fukugita:1988hg} and might also be due to a similar reduction of flux of  $\nu_e$. This was originally presented as a constraint on the GA parameter space in \cite{Conrad:2011ce}, but the shallowness of the respective $\Delta \chi^2$'s and the similarity in the parameter space leads us to consider the possibility of reconciling this data with the RAA and GA data.

Since the disappearance formulae are not sensitive to CP violation, the probability for both $\nu_e$ and $\bar{\nu}_e$ disappearance is given in Eq.~\eqref{disappsimple}. In all of the experiments in consideration, the neutrino energy is less than what we will find is the allowed range for $m_5$, so we set $a \to 0$ in all cases. It is clear from Eq.~\eqref{disappsimple} that $n_5$ can reduce the $\nu_e$ flux by a fixed amount with no energy or distance dependence whether or not $n_5$ is kinematically accessible. To extract the most conservative limits on the mixings with $n_5$ we will set $\sin^2 2 \theta_{e4}$ to 0 in these fits. We fit to all of the available data using correlation information as in the literature. We find a preferred value of
\beq \label{RGCval}
|U_{e5}|^2 = 0.036 \pm 0.013.
\eeq
Thus, we find that the RAA and GA may be consistently reconciled with the carbon data in the presence of a very heavy neutrino with averaged oscillations and a mixing angle of the magnitude indicated by Eq.~\eqref{RGCval}. However, we will show below that for a very massive sterile neutrino, a mixing angle of this magnitude is disfavored by a combination cosmological considerations, direct searches, and precision electroweak tests.

\subsection{Big Bang Nucleosynthesis}

Depending on the mixing and mass, additional light sterile neutrino(s) (with mass $\lesssim 1 \mev$) can be thermalized in the time leading up to big bang nucleosynthesis (BBN).  The presence of additional neutrinos at BBN can drive a faster expansion rate, modifying the abundance of the light elements, and in particular of helium.  Thus, detailed observations of primordial elemental abundances from BBN can constrain the properties of sterile neutrinos. To set bounds, we require that the total number of neutrinos at BBN is less than 4.4 \cite{Mangano:2011ar}, or $\Delta N_\nu \leq 1.4$.  With potentially two sterile neutrinos with masses below 1 MeV in the 3+1+1 scenario, BBN constraints must be carefully checked.  While constraints from BBN can be alleviated by the inclusion of a large lepton asymmetry (which effectively delays the time when an MSW-like coherent conversion can occur) \cite{leptonasymmetry}, this mechanism becomes ineffective for the large mass splittings of interest for the 3+1+1 model.  We review the constraints in this section and apply them to the 3+1+1 scenario.

We begin this discussion by reviewing the calculation for one active plus one sterile neutrino.  With this result in hand, we will be able to easily see how the result extends to two sterile neutrinos (with widely separated masses) mixed with more than one active neutrino.  We follow the density matrix formalism of  \cite{Dolgov:2003sg}.
Assuming that the active neutrinos are always in a fully thermalized state, the evolution equations for an arbitrary number of neutrinos $N$ is
\be
i \dot{\rho}=[\cH,\rho]-i\Gamma \rho,
\ee
where the Hamiltonian $\cH_{\alpha \beta}=V_{\alpha \beta}+\sum_{i=1}^N U_{\alpha i}U^*_{\beta i} m_i^2/2E$, and we take the production rate to be $\Gamma_{\alpha \beta}=(\Gamma_\alpha+\Gamma_\beta)/2$, with $V_{is}=\Gamma_s=0$ for all $i$.
Specializing to the case of one active and one sterile neutrinos, the relevant equations are
\begin{eqnarray}
H x \partial_x \rho_{ss} & = & i {\cal H}_{as}(\rho_{as}-\rho_{sa}) \\ \nonumber
H x \partial_x \rho_{as} & = & -i[({\cal H}_{aa} -{\cal H}_{ss}) - i \gamma_{as} ]\rho_{as} + i {\cal H}_{as}(\rho_{aa}-\rho_{ss}),
\end{eqnarray}
with $x = m/T$, and $m$ fixed to be $ 1\mev$.
The effect of interactions encapsulated in $\gamma_{as}$ is to damp away the coherent off-diagonal element, $\rho_{as}$.  Thus if $\gamma_{as}$ is large we are forced into the stationary point where $\partial_x \rho_{as} \approx 0$ \cite{Dolgov:2003sg}, so that
\be
\rho_{as} = \frac{{\cal H}_{as}}{({\cal H}_{aa} -{\cal H}_{ss}) - i \gamma_{as}}(\rho_{aa} - \rho_{ss}).
\ee
Substituting this in the differential equation for $\rho_{ss}$ we obtain
\be
H x \partial_x \rho_{ss} = \frac{\gamma_a}{4}(\rho_{aa} - \rho_{ss})\frac{\sin^2 2 \theta}{(\cos 2 \theta - V_{aa} / \delta E)^2 + \gamma_a^2/4\delta E^2},
\ee
where $\delta E = \Delta m^2/2 E$, $\gamma_a = g_a \frac{180 \zeta(3)}{7 \pi^4} G_F^2 T^4 p$ and $V_{aa} \simeq - C_a G_F^2 T^4 p/\alpha$, with  $g_{\nu_e} \simeq 4$, $g_{\nu_\mu,\nu_\tau} \simeq 2.9$, $C_{\nu_e} \simeq 0.61$, $C_{\nu_\mu,\nu_\tau} \simeq 0.61$.  If we neglect the $\gamma_a^2$ term in the denominator, which is valid for the non-resonance case, this is easily soluble analytically, since the result takes on the simple form:
\be
\ln(1-\Delta N_\nu) \approx \frac{\gamma_a(T = m)}{4 H(T = m)} \sin^2 2 \theta \int_0^1 {\rm d}x \frac{x^8}{(x^6\cos 2 \theta - V_{aa}(T = m)/\delta E(T = m))^2},
\ee
Doing the integral analytically or numerically, we see that the result scales as $\sim \sqrt{\delta E(T= m)/V_{aa}(T = m)} \sim \sqrt{\Delta m^2}$. The physical meaning of this result is clear.  The rate with which the sterile neutrino is populated is suppressed at late time because the interaction rate is dropping as $1/x^5$.  At the same time, the sterile neutrino is most likely to be populated when the mass splitting between the active and sterile states is smallest.  The medium dependent mass splitting, however, is also dropping with $V_{aa}$.  Altogether, the integral is dominated by when $x^6 \cos 2 \theta \sim |V_{aa}(T = m)/ \delta E(T = m)|$, so that the total result (squared) is \cite{Dolgov:2003sg}
\begin{eqnarray}
(\Delta m^2_{41}/\mbox{eV}^2) \sin^4 2 \theta_{es} &  = & 3.2 \times 10^{-5} \ln^2(1-\Delta N_\nu) \\ \nonumber
(\Delta m^2_{41}/\mbox{eV}^2) \sin^4 2 \theta_{\mu,\tau s} &  = & 1.7 \times 10^{-5} \ln^2(1-\Delta N_\nu).
\end{eqnarray}

With these physical insights, it is easy to see how the results generalize to the cases with more than one sterile or active neutrino.  First, we can see that because the sterile neutrino is populated around when $V_{aa}/\delta E \sim 1$, at any given temperature only one of the sterile neutrinos will be populated if the masses of the sterile neutrinos are widely separated in mass from each other and from the active neutrinos.  Thus, if we make the assumption that $m_a \ll m_4 \ll m_5$, we can decouple the fourth and fifth neutrinos from each other and treat them as being populated only through their interactions with the active neutrinos.

The other complication to consider is mixing between the active neutrinos themselves.  However, if the active neutrino mass splittings themselves are negligible in comparison to the sterile neutrino mass splittings, $\Delta m_{12}^2 \ll \Delta m_{23}^2 \ll \Delta m_{34}^2 \ll \Delta m_{45}^2$, then the SM mixing angles $\theta_{12},~\theta_{13},~\theta_{23}$ can be rotated away, and as a result the mixing between the active neutrinos itself decouples.  Thus we conclude that the constraints on active-sterile mixing can be decoupled accordingly, and we have
\begin{eqnarray}
(\Delta m^2_{(4,5)1}/\mbox{eV}^2) \sin^4 2 \theta_{e4,5} &  = & 3.2 \times 10^{-5} \ln^2(1-\Delta N_\nu) \\ \nonumber
(\Delta m^2_{(4,5)1}/\mbox{eV}^2) \sin^4 2 \theta_{\mu,\tau 4,5} &  = & 1.7 \times 10^{-5} \ln^2(1-\Delta N_\nu).
\label{activeAndsterile}
\end{eqnarray}

Now LSND and MB, in the standard 3+1 scenario, probe $\sin^2 2 \theta_{\mu e} = 4 |U_{e4}|^2 |U_{\mu 4}|^2 \simeq \frac{1}{4} \sin^2 2 \theta_{e4} \sin^2 2 \theta_{\mu 4}$, which is a good approximation in the small mixing angle limit.  Thus we are able to conclude that
\be \label{bbneq}
4 (\Delta m^2_{(4,5)1}/\mbox{eV}^2) \sin^2 2 \theta_{\mu e}   =  \sqrt{3.2 \times 1.7} \times 10^{-5} \ln^2(1-\Delta N_\nu),
\ee
so that we learn that the sterile neutrino is thermally populated if it has mixing angles large enough to explain LSND plus MB in the 3+1 scenario.

Now these results are easily extended to the 3+1+1 scenario.  Then we have constraints on  $r \sin^2 2 \theta_{\mu e} = 4 |U_{e4}^* U_{\mu 4}||U_{e4}^* U_{\mu 4}+U_{e5}^* U_{\mu 5}|$ from LSND plus MB.  To alleviate the constraints from the disappearance experiments, we require $|U_{e5}^* U_{\mu 5}| \gtrsim  |U_{e4}^* U_{\mu 4}|$.  Since $m_5 \gg m_4$, we conclude from Eq.~(\ref{activeAndsterile}) that if $\nu_4$ is populated, then $\nu_5$ is also populated at BBN temperatures, unless $m_5 \gtrsim 1 \mev$.  Since $\Delta N_\nu < 1.4$, we thus conclude that most of the $m_5$ parameter space proposed in \cite{Nelson:2010hz} is not consistent with the constraints from BBN, eliminating the entire region of parameter space with $33 \ev< m_5 \lesssim 1 \mev$.

Above the upper end of this mass range production of heavy sterile neutrinos may be inefficient at BBN temperatures, and the additional neutrino $n_5$ may not represent a fully populated degree of freedom. Because the state $n_4$ is fully populated, we require the fractional population of the state $n_5$ to satisfy $\Delta N_\nu^{(n_5)}\lesssim 0.4$ as calculated from Eq.~\eqref{bbneq}. After their production is frozen out, the remnant $n_5$ will decay through charge- and neutral-current interactions at rates suppressed by the mixing parameters. If these decays proceed through SM channels, the decays of $n_5$ can add entropy and ionizing energy to the thermal bath at the time of BBN, spoiling predictions of the relic helium abundance. These considerations allow us to rule out the range
\be
6.2 \times 10^{-10} \left( \frac{m_5}{\mev} \right)^{-1} \leq |U_{e5} U_{\mu 5}| \lesssim10^{1.5} \left(\frac{m_5}{\mev}\right)^{-3.5},
\ee
where the upper bound is a rough fit to the numerical analysis conducted in \cite{Ruchayskiy:2012si} (we display the numerical values in Fig.~\ref{meg}). These bounds extend up to $m_5 \simeq m_\pi$, at which point new decay channels open which have not been analyzed numerically.

In the next section we consider further constraints from colliders on such heavy neutrinos.  We will find that to satisfy the constraints, $n_5$ must have exotic invisible decays which are not via its SM mixing with the active neutrinos.

\subsection{Supernova 1987A}

The duration of the observed neutrino burst from Supernova 1987A (SN1987A) constrains the mass and couplings of any massive sterile neutrino. If the sterile neutrinos mix strongly enough that they are produced but are coupled weakly enough that they are not tightly bound to the supernova core they will allow too much energy to escape from the core, reducing the observed duration of the blast.

There are both lower and upper bounds on the neutrino coupling \cite{Kainulainen:1990bn}. The lower bound comes from requiring that $n_5$ are efficiently produced in the interior of the supernova. If these neutrinos are efficiently produced and have low enough mixing, they will free stream out of the supernova and conduct energy away from the core too quickly. With larger mixing angles, the neutrinos will have a short mean free path and, for large enough mixing, they will be trapped in the supernova. If they are trapped but their mean free path is larger than the supernova core they will cause anomalous cooling of the star: blackbody radiation will be emitted from a region larger than the supernova core, and the supernova will cool too quickly. Because production of neutrinos in supernovae are dominated by charge-current processes, we find that $\nu_e$ production dominates $\nu_\mu$ production \cite{Raffelt:2003en}. This gives slightly weaker bounds on the mixing angle $U_{\mu5}$, and, because maximal mixing angles are in principle allowed by these arguments, we find that the lower bound on $U_{\mu5}$ (which is approximately 5 times weaker than the lower bound on $U_{e 5}$) is in fact the lower bound on the product $|U_{e5} U_{\mu5}|$. For $m_5 \gtrsim 0.1\mev$, where the matter effect becomes unimportant, we find that mixing angles $3.0 \times 10^{-5} \lesssim |U_{e5} U_{\mu 5}| \lesssim 5.0 \times 10^{-3}$ are ruled out by these energy considerations. These bounds apply to $n_5$ regardless of its couplings.

The trapping argument given above will not apply for large mixing angles if $n_5$ decays invisibly to products with no SM interactions, as naturally considered in the model building section below. This is because, for the widths calculated below, we find that the decay length $L \simeq 10^{-10} {\rm ~m~} (10 \mev/m_5)$ is much shorter than the mean free path $\lambda_{\rm mfp} \simeq 0.1 {\rm ~m~} /\sin^2 2 \theta_m$ for the masses and mixings of interest. Therefore $n_5$ will decay well before it is trapped, and the bounds can no longer be lifted at very large mixing angles. These exclusions are model-dependent because they rely on the unknown couplings of the decay products, but since it is possible that $n_5$ evades the upper bounds described at large mixing angles we shade this region gray in the left panel of Fig.~\ref{meg}.

These bounds are also lifted for lighter sterile neutrinos since the relevant production mechanism is matter-enhanced flavor transitions. For $m \lesssim \cO(0.1 \mev)$ the bounds weaken and go to zero around $m \sim \cO(100 \ev)$ \cite{Kainulainen:1990bn}, so SN1987A bounds do not constrain $n_4$.

\subsection{Bounds from Light Mesons}

There are a variety of searches for exotic meson decays that produce strong bounds on the mass and mixing of $n_5$. We group these into a few categories as follows and display the collected results in Fig.~\ref{meg}.
\\ \\
\noindent  {\bf Measured $\pi$ meson branching fraction:}
The pion branching ratio to $\mu$ and $e$ is $R_{\pi}=|\overline{\cM}|_{\pi \to e \nu}^2/|\overline{\cM}|_{\pi \to \mu \nu}^2$.  At tree level, the matrix element $|\overline{\cM}|_{\pi \to \ell_\alpha \nu}^2$ goes like
\be
|\overline{\cM}|_{\pi \to \ell_\alpha n}^2 \propto \sum_{i=1}^{N_a} |U_{\alpha i}|^2  \left( m_\alpha^2+m_i^2 \right) \left[m_{\pi}^2-(m_\alpha+m_i)^2 \right], \label{pionth}
\ee
where $m_\alpha$ is the mass of the charged lepton $\ell_\alpha$. In the SM, where $m_i=0$ for all neutrinos and $U_{\ell_\alpha i}=\delta_{\ell_\alpha i}$, $R_{\pi}$ simplifies considerably. The most current SM calculation of this quantity to two loops is $R_{\pi}^{\rm SM,th.}=(1.2352 \pm 0.0001) \times 10^{-4}$ \cite{Cirigliano:2007xi}, while the best experimental bounds give $R_\pi=(1.230 \pm 0.004) \times 10^{-4}$ \cite{PDG}.

In the 3+1+1 framework $R_\pi$ will differ depending on the mass range, so the constraints are piecewise. They simplify at high mass, where $m_5 \gtrsim m_\pi - m_\mu \gg m_e$, which is near where the SN1987A bounds stop. We find
\be
\frac{R_\pi}{R_\pi^{\rm SM,th.}} \simeq \left\{ \begin{array}{cl}
\frac{1-|U_{e5}|^2}{1-|U_{\mu 5}|^2} + \frac{|U_{e5}|^2}{1-|U_{\mu 5}|^2} \frac{m_5^2}{m_e^2} \frac{m_\pi^2-m_5^2}{m_\pi^2-m_e^2} & m_\pi-m_\mu \lesssim m_5 \lesssim m_\pi \\ \frac{1-|U_{e5}|^2}{1-|U_{\mu 5}|^2} & m_5 \gtrsim m_\pi \end{array} \right.
\ee
The measured ratio is $R_\pi/R_\pi^{\rm SM,th.}=0.996 \pm 0.003$, so at 99\% confidence we require that $R_\pi/R_\pi^{\rm SM,th.} \lesssim 1.004$. Thus, the mixing angles $U_{e5}$ and $U_{\mu 5}$ are bounded fairly strongly in the intermediate mass range. We do a scan over the full parameter space and for each value of $m_5$ we find the maximum product of the mixing angles consistent with this constraint.
\\ \\
\noindent  {\bf Muon lifetime:}
For nonzero $U_{e5}$ and $U_{\mu5}$, the total charged current interactions with the muon and electron below the muon mass will be reduced. The muon lifetime $\tau_\mu$ will be increased relative to the SM prediction due to the non-unitarity in the neutrino mixing matrix. In practice, Fermi's constant, $G_F$, is measured most precisely from measurements of $\tau_\mu$ \cite{Barczyk:2007hp}, so one can derive constraints by comparing to an independent measurement of $G_F$. Following \cite{Biggio:2009nt}, we relate $M_Z$, $M_W$ and $\alpha$ to $G_F$ by
\be
G_F^\prime = \frac{\pi \alpha M_Z^2}{\sqrt{2} M_W^2 (M_Z^2 - M_W^2)(1-\Delta r)}.
\ee
where $\Delta r = 0.0362\pm 0.0005$ \cite{PDG} is the correction to the tree-level relationship. The values of $M_Z$ and $M_W$ used should be taken from purely kinematic measurements since other fits to $M_W$ include the measurements of $G_F$ from muon decay. We take $M_W = 80.387 \pm 0.016$ GeV \cite{TevatronElectroweakWorkingGroup:2012gb} and  $M_Z = 91.1875 \pm 0.0021$ GeV \cite{ALEPH:2005ab}. Plugging in these values, we find $G_F^\prime = (1.1679 \pm 0.0013) \times 10^{-5} \gev^{-2}$. Comparing this to the value extracted from measurements of $\tau_\mu$, $G_F= 1.166 353(9)\times 10^{-5} \gev^{-2}$ \cite{Barczyk:2007hp}, we find for $m_5 > m_\mu$
\be
\frac{G_F}{G_F^\prime} = (1-|U_{e5}|^2)(1-|U_{\mu5}|^2) = 0.9987 \pm 0.0011,
\ee
resulting in an upper limit
\be
|U_{e5} U_{\mu5}| < 0.0021
\ee
at 99\% CL. We mark this line as $\tau_\mu$.
\\ \\
\noindent {\bf Searches for lines in $\pi$ and $K$ meson decays:} Measurements of $\pi,K \to \ell_\alpha n$ give important bounds on the mass and mixing of $n_5$ with $\nu_\alpha$.  These are summarized in Figs.~(2-4) of \cite{Atre:2009rg}. Bounds are given by null searches for  peaks in the spectra of the leptonic products of these decays. For $n_5$ produced by the decay of a heavy parent particle $M$ of mass $m_M$ with a decay partner $\ell_\alpha$ of mass $m_\alpha$ we expect to see a monochromatic line in the lepton spectrum at $E_\alpha=\left( m_M^2 +m_\alpha^2-m_5^2 \right)/2m_M$. These lines are generically not found, and limits on $n_5$ mixing are based on the specifics of the given experiment.

For the electron neutrino sector, the decay $\pi \to e \nu$ \cite{Britton:1992xv} is strongest below $m_\pi$ and $K \to e \nu$ \cite{Yamazaki:1984sj} is strongest between $m_\pi$ and $m_K$. For the muon neutrino sector, the important decay is $K \to \mu \nu$ \cite{Hayano:1982wu}. In the region $m_5>m_\pi-m_\mu$, muons cannot be produced in $\pi$ decay so there are no muon bounds in that range. Thus, line searches of $\pi$ decays do not provide strong constraints on the product $| U_{e 5} U_{\mu 5}|$ in the mass range $m_5>m_\pi-m_\mu$ since experiments cannot set any bounds on $U_{\mu 5}$ in this range.
\\ \\
\noindent {\bf Decays of $n_5$:} If $n_5$ is heavy and can decay to SM products, these decays will be seen in dedicated searches such as, {\it e.g.}, the PS191 \cite{Bernardi:1987ek}, CHARM \cite{Bergsma:1985is}, and DELPHI \cite{Abreu:1996pa} experiments. PS191 looked for the decay of a heavy neutrino through a variety of weak interaction channels; it is constraining from $\sim 1\mev$\footnote{The PS191 experiment did not publish limits for mixing angles above $10^{-4}$, so we extrapolate the bounds down to $m_5=2m_e$ as a power law with $\propto \sqrt{m_5^5}$.} to 138 MeV. CHARM searched for decays $n_5 \to \ell^+ \ell^- \nu$, where $\ell=e,\mu$, with constraints from 500 MeV to 2.8 GeV. DELPHI also looked for a wide variety of $n_5$ decays, and it provides limits from 2 GeV to 90 GeV. We show these excluded regions as well as the limits from $n_5$ decays in dileptonic $K$ decays \cite{PDG}. Note that, as pointed out previously \cite{Kusenko:2004qc}, the PS191 and CHARM collaborations considered $n_5$ decays through charge-current channels only. When the necessary neutral-current contributions are added \cite{Atre:2009rg}, the bounds are strengthened somewhat compared to the published results \cite{Kusenko:2004qc}. We provide bounds including both the charge- and neutral-current contributions.
\\ \\
\noindent {\bf Non-observation of $\mu \to e \gamma$:}
For the decay $\mu \to e \gamma$, we have the standard result
\beq
\Br(\mu \to e \gamma)=\frac{3 \alpha}{8 \pi} \left| \sum_i U_{ei} U^*_{\mu i} ~g (m_i) \right|^2,
\eeq
where $g(m_i)$ is a kinematic factor given in  \cite{Atre:2009rg}. This is not constraining below $\cO$(1 GeV) and by 300 GeV the bound asymptotes to $\left| U_{e5} U_{\mu 5}  \right| \lesssim 5.25 \times 10^{-5}$ using the current measurement $\Br(\mu \to e \gamma) \leq 2.4 \times 10^{-12}$ \cite{Adam:2011ch}. At high mass, this is the most important constraint. In particular, measurements at the $Z$-pole are weaker than $\mu \to e \gamma$, so we do not show these bounds on our plots.

\subsection{Combined Bounds on $n_5$}

\begin{figure}[t]
\begin{center}
\includegraphics[width=6.6in]{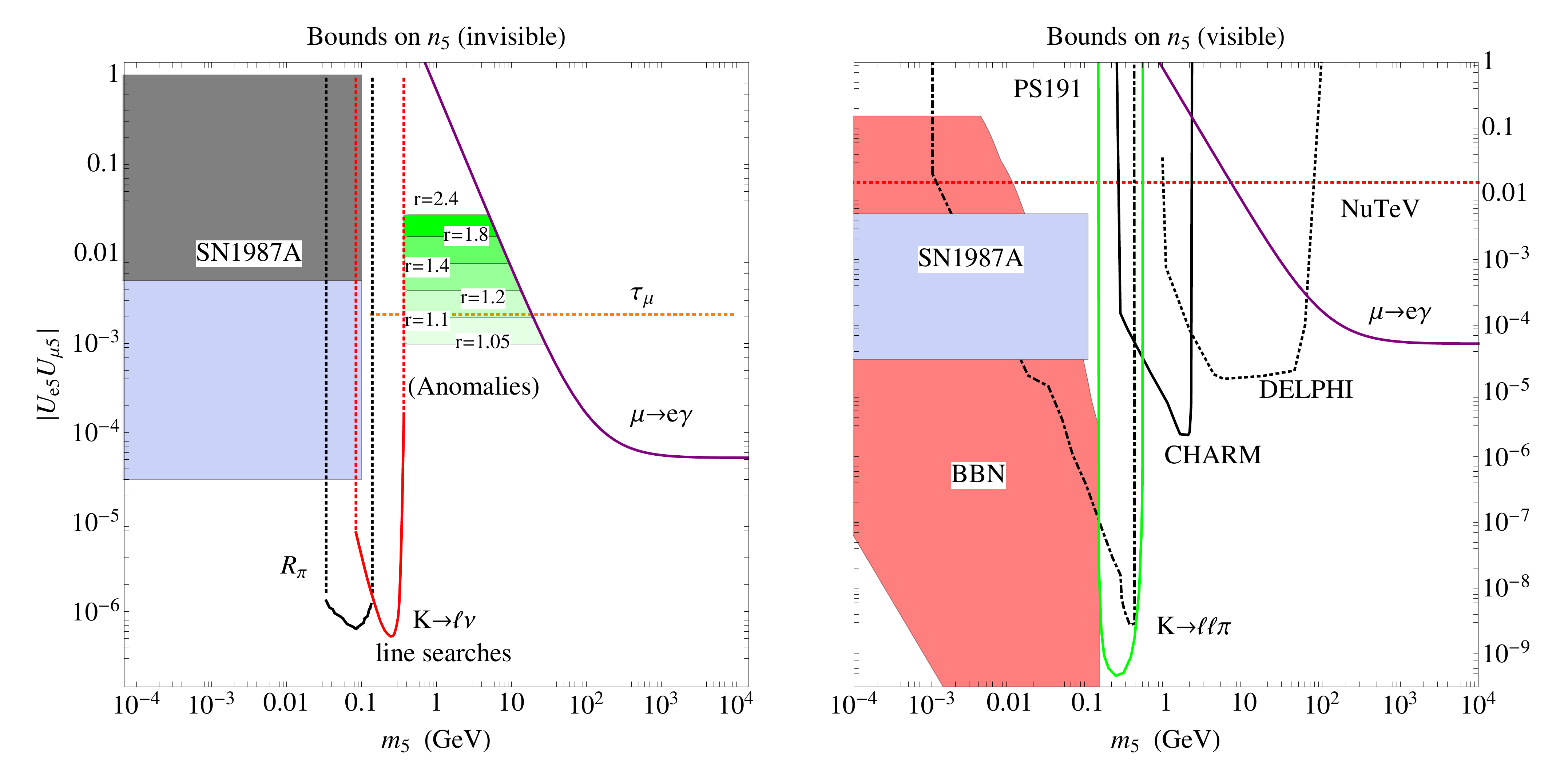}
\caption{Exclusion regions from BBN \cite{Ruchayskiy:2012si} (right frame) and SN1987A \cite{Kainulainen:1990bn} (both frames), as well as bounds from the NuTeV oscillation search \cite{Avvakumov:2002jj} (red dotted, right frame), $R_\pi$ \cite{Cirigliano:2007xi,PDG} (left frame), measurements of $\tau_\mu$ \cite{PDG,Barczyk:2007hp,Biggio:2009nt,TevatronElectroweakWorkingGroup:2012gb,ALEPH:2005ab} (left frame),  collider and line searches \cite{PDG,Britton:1992xv,Yamazaki:1984sj,Hayano:1982wu,Bernardi:1987ek,Bergsma:1985is,Abreu:1996pa,Kusenko:2004qc} (both frames), and searches for $\mu \to e \gamma$ \cite{Adam:2011ch} (both frames). The left panel shows lines of constant values of $r$ from 1.05 to 2.4 (for the calculation of $r$, we assume no CP violation and take $|U_{e4} U_{\mu 4}| = 0.023$, as explained in the text). To avoid clutter, we avoid repeating the $\tau_\mu$ and NuTeV lines in both plots, although each is valid in both cases.}
\label{meg}
\end{center}
\end{figure}

In Fig.~\ref{meg}, we show bounds on the product $\left| U_{e5} U_{\mu 5} \right|$, which in the CP-conserving limit is the product that sets the value of $r$ in the appearance probability formula, Eq.~\eqref{appmaster}. There are several model-independent bounds: as described above, BBN is most constraining below $\sim1 \mev$; there are universally constraining bounds for masses $0.1~\mev<m_5<100~\mev$ from SN1987A;  the NuTeV oscillation search \cite{Avvakumov:2002jj} rules out mixing angles $|U_{\mu 5}  U_{e 5}|>1.3 \times 10^{-2}$ for the entire mass range; and for masses $m_5>64\gev$, the bounds from $\mu \to e \gamma$ are most stringent. In the range 100 MeV $<m_5<64\gev$ the constraints bifurcate depending on whether $n_5$ decays to charged leptons or remains invisible on collider timescales.
\\ \\
{\bf Invisible decays:} When $n_5$ remains invisible on collider timescales there are constraints from line searches, the pion branching fraction, and precision electroweak measurements of $G_F$. From $m_\pi-m_\mu<m_5<m_\pi$ the strongest bound is from the measured branching fraction of pion decays, $R_\pi$. For $m_\pi < m_5< m_K-m_\pi$ searches for leptonic lines in kaon decays are constraining for both $e$ and $\mu$ products. For $m_\mu < m_5$, comparing the values of $G_F$ from measurements of $\tau_\mu$ and the $W$ and $Z$ masses as described above gives tight constraints. We show these bounds in the left panel of Fig.~\ref{meg}.

Between the $K$ line searches and the $\mu \to e \gamma$ curve, where $ 387 \mev<m_5 \lesssim 10\gev$, the most constraining bounds on $n_5$ come from the precision electroweak measurements of $G_F$. Although this is the least constrained region, we find that these measurements still disfavor large values of $r$. Assuming no CP violation and taking $|U_{e4} U_{\mu 4}| = 0.023$, which is the smallest value of $|U_{e4} U_{\mu 4}|$ for which $|U_{e5} U_{\mu 5}|$ can take on arbitrarily low values in the MB and LSND region, we find that $r < 1.09$.
\\ \\
{\bf Visible decays:} In addition to the SN1987A and $\mu \to e \gamma$ bounds and the low mass constraints on BBN, we find that the direct searches at PS191, CHARM, and DELPHI are very constraining if $n_5$ decays to SM particles on detector timescales, and we also find that the BBN constraints can be extended to $m_5 \simeq m_\pi$, as described above. These give the most powerful constraints from $\lesssim 1 \mev$ to 64 GeV. Above this range, the $\mu \to e \gamma$ constraints become powerful. These bounds are in the right panel of Fig.~\ref{meg}.
~
~
\\ \\
We see in Fig.~\ref{meg} that the bounds are prohibitively strong if $n_5$ decays to SM products. We find that $n_5$ is phenomenologically more viable provided the decays of $n_5$ are invisible and the mass satisfies $387\mev<m_5 \lesssim 10\gev$. However, when restricting the range of $|U_{e5} U_{\mu 5}|$ from the muon lifetime, the LSND and MB results strongly favor $r\sim1$.   Furthermore, the combination of constraints from  SN1987A and the muon lifetime restricts $|U_{e5}|^2 < 0.004$ for $m_5 \gtrsim 100\mev$, which seriously constrains the parameter space for solving the RAA and GA data, as indicated by Eq.~\eqref{RGCval}.

In the next section we construct models of neutrino mass that naturally allow for invisible decays.

\section{Neutrino Models}

The 3+1+1 scenario relies on the presence of a heavy neutrino with a substantial mixing with the light neutrinos.  Within the standard see-saw scenario, with one active neutrino $\nu_a$ and one sterile neutrino $\nu_m$, this is not possible to achieve.  The mass matrix
\beq
\cM=\left( \begin{array}{cc}
0 & m_D \\
m_D& M
\end{array}\right)
\eeq
connects the mixing to the mass hierarchy, so that a heavy sterile neutrino necessarily has a small mixing with the SM neutrino: $\theta \sim m_D / M, $ which is small for a sizable neutrino hierarchy.

A large mass hierarchy and a large mixing can, however, be achieved for a Dirac sterile neutrino.  Taking a single active neutrino $\nu_a$ and a sterile neutrino $\nu_d$ with Dirac partner $\bar{\nu}_d$, we can write a general mass matrix in the $(\nu_a,\nu_d,\bar{\nu}_d)$ basis as
\beq
\cM=\left( \begin{array}{ccc}
m_1 & m_D & 0\\
m_D& 0 & m_5 \\
0 & m_5 & 0
\end{array}\right).
\eeq
Defining $M^2 \equiv m_5^2+m_D^2$ and expanding to second order in the small ratio $m_1/m_5$ we find the eigenvalues
\begin{align}
\lambda_1&=M+\left( \frac{m_D^2}{2M^2} \right) m_1 + \left[ \frac{m_D^2 \left( m_D^2 + 4m_5^2 \right)}{8 M^5} \right] m_1^2 + \dots \nonumber \\
\lambda_2&=-M+\left( \frac{m_D^2}{2M^2} \right) m_1 - \left[ \frac{m_D^2 \left( m_D^2 + 4m_5^2 \right)}{8 M^5} \right] m_1^2 + \dots\\
\lambda_3&=\frac{m_5^2}{M^2}m_1 \nonumber,
\end{align}
corresponding to the (unnormalized) eigenvectors
\beq
K_1=\left( \begin{array}{c} \frac{m_D}{M} \left( 1+\frac{m_1}{M} \right) \\ 1 \\  \frac{m_5}{M} \end{array} \right) \qquad
K_2=\left( \begin{array}{c} -\frac{m_D}{M} \left( 1- \frac{m_1}{M} \right) \\ 1 \\  - \frac{m_5}{M} \end{array} \right) \qquad
K_3=\left( \begin{array}{c} -\frac{M^2}{m_1m_D} \\ 1 \\  \frac{M^2}{m_1m_5} \end{array} \right) .
\eeq
$K_1$ and $K_2$ correspond to the components of the mostly sterile fifth mass eigenstate $n_5$, whereas $K_3$ corresponds to a mostly active light state.  The mixing between $n_5$ and the light state is controlled by $m_D/M$. This ratio need not be very small since the mass of the light neutrino is fixed independently by $m_1$. The small mixing scenario is recovered in the limit $m_D \ll m_5$, which corresponds to $m_5 \to M$, while maximal mixing corresponds to the limit $m_D \to m_5$.

This type of scenario can be extended to encompass the fourth neutrino, as well as the needed invisible decays of $n_5$.  Consider adding to the Lagrangian a term
\be
{\cal L}_\phi = \lambda \phi \nu_d \nu_m + \lambda' \phi \nu_m^2.
\ee
Neglecting Majorana mass terms for illustration, we find that a mass matrix in the $(\nu_a,\nu_d,\bar{\nu}_d,\nu_m)$ basis with the desired phenomenology is given by
\beq
\cM=\left( \begin{array}{cccc}
0 & m_D & 0&0\\
m_D& 0 & m_5&m_\phi \\
0 & m_5 & 0&0 \\
0&m_\phi&0&0
\end{array}\right),
\eeq
where $m_\phi = \lambda v_\phi$.  This matrix has two zero eigenvalues, with the other two set by $\pm \sqrt{m_\phi^2+m_D^2+m_5^2}$.  In a hierarchy where $m_\phi \ll m_D \ll m_5$, the massive states are mostly $\nu_d$ and their mixing with the active neutrino is controlled by $m_D/m_5$.  The massless states are predominantly composed of $\nu_a$ and $\nu_m$, and their mixing is controlled by $m_\phi/m_D$. Of course, these masses should not exactly vanish, and the masses can be lifted  from being zero by appropriately small Majorana mass terms.

The new state $\phi$ allows both for large $\nu_d-\nu_a$ mixing and for invisible decays of $\nu_d$ (via $\nu_d \rightarrow \phi \nu_m$ with subsequent decays $\phi \rightarrow \nu_m \nu_m$).  This decay, with width $\Gamma_\phi \sim \frac{1}{16\pi}\lambda^2 m_5$, should be compared to the decay derived from mixing with active states, which scales as $\Gamma_{SM} \sim \frac{1}{16 \pi}\theta_{\mu,e}^2 g_Z^2 \left(\frac{m_5}{m_Z}\right)^4 m_5$.  Since $m_5$ is in the GeV range, the SM decay channel is naturally suppressed with respect to the invisible decay.

\section{Conclusions}

We have studied fits to the LSND and MB experiments within the context of 3+1 and 3+1+1~\cite{Nelson:2010hz} scenarios.  Compared to the 3+1 scenario, the 3+1+1 framework posits that the presence of an additional heavy neutrino which is not directly probed by most disappearance experiments lifts some of the constraints of the null disappearance experiments.  However, using the new 2011 MB $\bar{\nu}$ data, we find there is still significant tension between positive and the null results, even with the additional very heavy neutrino.  We went on to explore the phenomenology of the massive neutrino that appears in the 3+1+1 scenario, and we found that a heavy mostly sterile neutrino could be consistent with a variety of cosmological and collider constraints if the sterile neutrino has a mass around a GeV and does not couple primarily to the SM. We also showed that in the face of BBN, direct search, and precision electroweak bounds, even a heavy state that decays invisibly might not be suitable for reconciling the anomalous measurements of $\nu_e$ fluxes made by gallium and reactor experiments.

If the larger mixing angle required by the 2011 $\bar{\nu}$ data for the LSND and MB anomaly persists, other types of scenarios will be required in order to obtain a consistent global explanation of the neutrino oscillation data.  One possibility, which was explored in \cite{Zurek:2004vd}, is to make use of medium dependent neutrino masses \cite{Kaplan:2004dq}.  In this case, Bugey (whose oscillations would mostly occur through air) would be weakened relative to LSND and MB (whose oscillations mostly occur through earth), and a wide swath of parameter space would remain.  We leave this possibility for future consideration.

{\em Acknowledgments:} We thank Ann Nelson and Tomer Volansky for discussions. We would also like to thank Enrique Fernandez Martinez, Oleg Ruchayskiy, and Jinrui Huang for pointing out some additional bounds relevant to the parameter space of Fig.~\ref{meg}. The work of EK is supported in part by a grant from the Israel Science Foundation.  The work of KMZ and SDM is supported by NSF CAREER PHY 1049896, and NASA Astrophysics Theory Program Grant No. NNX11AI17G.

\appendix

\section{Fit Details}
\noindent We summarize and expand upon the details of our analysis of experimental data included in our fits.
\\
\\
\noindent {\bf Appearance Experiments:}
The characteristics of the experiments used in our fits are shown in Table~\ref{apptab}.  LSND \cite{hep-ex/0104049} observed the appearance of $\bar{\nu}_e$ with energies $10-60$ MeV in a beam of $\bar{\nu}_\mu$, consistent with neutrino oscillations that occur in the $\Delta m^2 \sim 0.2 - 10 \ev^2$ range.  MB also measured $\bar{\nu}_\mu \to \bar{\nu}_e$ oscillations \cite{AguilarArevalo:2010wv} with energies $200-3000$ MeV consistent with evidence for antineutrino oscillations from LSND. MB did not initially report evidence for oscillations of the form $\nu_\mu \to \nu_e$ \cite{AguilarArevalo:2007it}, but an in-depth analysis published after the release of the initial data set supported the interpretation of a low-energy excess consistent with $\nu_\mu$ oscillations \cite{AguilarArevalo:2008rc}. We use all of the MB $\nu$ data points in our analysis, including those below 375 GeV which were excluded in the first MB analysis. Due to the low energies of these experiments, we take $a=0$ for our fits.

We also include the null results of KARMEN \cite{Armbruster:2002mp}, E776 \cite{Borodovsky:1992pn},  NOMAD \cite{Astier:2003gs}, CCFR \cite{Romosan:1996nh}  and NuTeV \cite{Avvakumov:2002jj}. For each null experiment, we find a single data point--that is, the oscillation probability and error for the $L/E$ value--which best matches the 90\% exclusion curves given by the experiments.  Due to the generally high energies of these experiments, we use $a=1$ for all experiments except KARMEN in these fits.

\begin{table}[h]
$$ \begin{array}{ l | c | c | c | c | c }  {\rm Experiment} & {\rm mode} & {\rm \# ~points} &  \text{Distance}~({\rm m})  & E & \Delta m^2~( {\rm eV^2}) \\ \hline  \hline
{\rm MB} & \bar{\nu}_\mu , ~\nu_\mu & 11\times 2 & 541  & 200-3000 \mev & \gtrsim 0.1 \\ \hline
{\rm LSND} & \bar{\nu}_\mu & 8 & 29.8  & 10- 60\mev & \gtrsim  0.3  \\ \hline
{\rm KARMEN} & \bar{\nu}_\mu & 1 & 17.7  & 1-50 \mev & \gtrsim 1  \\ \hline
{\rm E776} & \bar{\nu}_\mu,~{\nu}_\mu & 1 & 1000  & 1-10 \gev & \gtrsim 1 \ \\ \hline
{\rm NOMAD} & \nu_\mu & 1 & 625  & \gtrsim  10-200 \gev & \gtrsim 10  \\ \hline
{\rm  NuTeV} & \bar{\nu}_\mu,~{\nu}_\mu & 1 & 1436  & \gtrsim 10-300 \gev & \gtrsim 10^2  \\ \hline
{\rm CCFR} & \nu_\mu & 1 & 1436 & \gtrsim 10-300 \gev & \gtrsim 10^2  \\ \hline  \hline
{\rm TOTAL} & \bar{\nu}_\mu, \nu_\mu & 30{\rm~pos.,~5~null} & \sim 10 -1436 & 10\mev -~ 600\gev & \gtrsim  0.1  \\ \hline
\end{array}$$
 \caption{Energies, mixings, and mass splitting sensitivities for each appearance experiment.  \label{apptab}}
 \end{table}
 ~
\\
\\
\noindent {\bf Disappearance Experiments:}
The $\nu_e$ disappearance constraints include short-baseline reactor experiments with new reactor flux predictions \cite{Mention:2011rk} plus constraints from the ratio of the flux observed in the Bugey 40 m and 15 m detectors \cite{Declais:1994su}.\footnote{Even though the new reactor flux has been reported to reflect oscillations of a single sterile neutrino, we find that it is not consistent with our LSND and MB preferred region for the light sterile neutrino, and we use the reactor data as a constraint.  On the other hand, it may be fit well with the fifth neutrino of the 3+1+1 scenario, as shown in Sec.~\ref{sec:RGC}.} The statistics of the constraint on $\nu_e$ disappearance is dominated by the Bugey ratio.  Disappearance of $\nu_\mu$ is constrained by CDHS \cite{Dydak:1983zq} and CCFR \cite{Stockdale:1984ce}, which we take as single data points, corresponding to the combined oscillation probability for the full energy range. Because both of these experiments search for muon neutrino oscillations between two detectors, very large mass differences are not restricted, since the beam is likely to be fully oscillated as it arrives at both the near and the far detector. Also, because of the baselines and energies of these experiments, it is a good approximation to ignore the probability of oscillation through $n_{1,2,3}$. We take $a=1$ for CCFR and CDHS, but have $a=0$ for the reactor experiments.

\begin{table}[h]
$$ \begin{array}{ l | c | c | c | c | c }  {\rm Experiment} & {\rm mode} & {\rm \# ~points} &  \text{Distance} ~({\rm m}) & E & \Delta m^2~( {\rm eV^2}) \\ \hline  \hline
{\rm CCFR} & \nu_\mu & 1 & 714 {\rm~and~}1116 & 40-200 \gev & 10-10^3 \\ \hline
{\rm CDHS} & \nu_\mu & 1 & 130 {\rm~and~}885 & 2-6 \gev & 10^{-1}-10 \\ \hline
{\rm Mention}~et~al. & \bar{\nu}_e & 21 & 9-1050 & \sim 3 \mev & 10^{-2}-10^{-1} \\ \hline
{\rm Bugey~40/15~ratio} & \bar{\nu}_e & 25 & 15 {\rm~and~}40 & 3-8 \mev & \gtrsim 10^{-2} \\ \hline \hline
{\rm TOTAL} & \bar{\nu}_e, \nu_\mu & 48 & 10-10^3 & 3{\rm~MeV}-200 \gev & 10^{-4}-10^3 \\ \hline
\end{array}$$
 \caption{Energies, mixings, and mass splitting sensitivities for each disappearance experiment.  \label{distab}}
 \end{table}
 ~
\\
\\

\noindent {\bf Unitarity Constraints:} The condition for unitarity in the neutrino mass mixing matrix is $U^\dagger U=1$, or in component form $\sum_i U_{\alpha i} U^*_{\beta i} = \delta_{\alpha \beta}$. In practice, this means that sum of the norms of any single row or column in the mixing matrix must equal 1, which bounds the size of any particular element of the matrix.  In this way, high confidence measurements of mixing angles for $\nu_e$ and $\nu_\mu$ with $n_1$, $n_2$, and $n_3$ can set bounds on the size of the mixings of $\nu_e$ and $\nu_\mu$ with $n_4$ and $n_5$.

For instance, solar neutrino experiments such as KamLAND \cite{:2008ee} measure $\nu_e$ disappearance via the mixing $\sin^2 2 \theta_{sol}= 4 \left|U_{e 2}\right|^2 \left( 1- \left|U_{e 2}\right|^2 - \left|U_{e 3}\right|^2 - \left|U_{e 4}\right|^2 -   \left|U_{e 5}\right|^2 \right)$. We can extremize this over the mixing $|U_{e2}|^2$, but we find that the limits on $\left|U_{e 4}\right|^2 + \left|U_{e 5}\right|^2$ are not very constraining because the solar mixing angle is measured at low confidence and the mixing angle is not maximal. The limits from reactor $\bar{\nu}_e$ disappearance experiments provide stronger limits.

On the other hand, a similar analysis is effective in constraining $\left|U_{\mu 4}\right|^2 $ and  $ \left|U_{\mu 5}\right|^2$ from the Super-Kamiokande data, since the atmospheric mixing angle is measured at high confidence to be maximal: a substantial mixing with heavy sterile neutrinos would imply a larger than observed ratio in upward to downward going muon neutrino fluxes.  This mixing angle is $\sin^2 2\theta_{atm}= 4 \left|U_{\mu3}\right|^2 \left( 1-  \left|U_{\mu3}\right|^2 - \left|U_{\mu 4}\right|^2 -   \left|U_{\mu 5}\right|^2 \right)$, and extremizing to find the largest value of $\left|U_{\mu 4}\right|^2 + \left|U_{\mu 5}\right|^2$ compatible with the measurements gives
\beq
\sin^2 2\theta_{atm} \le \left( 1- \left|U_{\mu 4}\right|^2 - \left|U_{\mu 5}\right|^2 \right)^2.
\eeq
Using the global best fit value for atmospheric mixing angle we find the 90\% (99\%) confidence level constraints:
\be
\left|U_{\mu 4}\right|^2 +   \left|U_{\mu 5}\right|^2 < 0.0175 ~(0.0274).
\ee
These bounds on $\left|U_{\mu 4}\right|^2 + \left|U_{\mu 5}\right|^2$ are included in the disappearance constraints, and are in practice the strongest constraints available.
 ~
\\
\\

\noindent {\bf Reactor and Gallium Anomalies:}
The RAA is detailed in \cite{Mention:2011rk,Huber:2011wv} and corresponds to a lower-than-expected flux of $\bar{\nu}_e$ emitted from nuclear reactors. The GA has been reported in \cite{gallium:ANOMALY} and also discussed clearly in \cite{Mention:2011rk} and \cite{Giunti:2010zu}: SAGE and Gallex have independently measured a lower-than-expected flux of $\nu_e$ from the decay of megacurie sources of ${}^{51}$Cr and ${}^{37}$Ar, corresponding to anomalously low rates of the reaction $\nu_e+{}^{71}$Ga$  \to {}^{71}$Ge $+~ e^-$. This is in principle bounded by similar measurements of the rate of $\nu_e +{}^{12}{\rm C \to {}^{12}N_{g.s.}} + e^-$ \cite{carbon:KARMENLSND,Conrad:2011ce}, which are more consistent with expectations \cite{Fukugita:1988hg}.  We find that these carbon data are compatible with fits to the RAA and GA data.

We summarize the status of the RAA and GA anomalies as well as the carbon data in Table~\ref{GCtab}. In Eq.~\eqref{RGCval} we give our fit to the combined data (using correlation information as reported in \cite{Mention:2011rk}).

\begin{table}[h]
$$ \begin{array}{ l | c | c | c }  {\rm Anomaly} & {\rm \# ~points} & |U_{e5}|^2  & \chi^2_{\rm min} \\ \hline  \hline
{\rm Gallium} & 4 & 0.0708 \pm 0.0317 & 1.7 \\ \hline
{\rm Carbon} & 11 & 0.0901 \pm 0.0874  & 8.5 \\ \hline
{\rm Reactor} & 19 & 0.0266 \pm 0.0144 & 7.2 \\ \hline \hline
{\rm TOTAL} &  34 & 0.0356 \pm 0.0130 & 19.4 \\ \hline
\end{array}$$
\caption{Fits to the reactor, gallium, and carbon anomalies.} \label{GCtab}
\end{table}
 ~
\\
\\
\noindent {\bf Statistical Methods Employed:}
To place constraints on neutrino mixing from the various null appearance and disappearance experiments we use the raster-scan method described in \cite{Feldman:1997qc}. For each value of $\Delta m^2_{41}$, the $\chi^2$ is minimized  with respect to the remaining mixing parameters. For the $n$ remaining independent mixing parameters, a $\Delta \chi^2$ test is performed to give an $n$-dimensional confidence interval at each $\Delta m^2_{41}$. In the 3+1 scenario, this corresponds to finding 1-dimensional confidence interval for $\sin^2 2 \theta$ at each given value of $\Delta m^2_{41}$.
The raster-scan provides a more precise confidence region than  a global fit. As a result of the sinusoidal dependence of the oscillation probability on $\Delta m^2_{41}$, the probability distribution for the $\chi^2$ may deviate from gaussian for large deviations from the true value. This may result both in finding an ``incorrect'' minimum of the $\chi^2$ and using an incorrect probability distribution function for determining the sizes of the confidence intervals.  By performing a raster scan, one removes the sinusoidal dependence so that the data follow a standard $\chi^2$ distribution.

Although this a powerful technique for forming exclusions from null experiments, it is less applicable to cases in which there is a positive result. This is because the raster-scan does not identify preferred values of the parameter $\Delta m^2_{41}$. For this reason we perform a global fit to the LSND and MB data, minimizing the $\chi^2$ with respect to all parameters. The confidence region is given by a $\Delta \chi^2$ for 2 DOF in 3+1 model  and 4 DOF in the 3+1+1 model.

\section{Oscillation Formalism}

For nonrelativistic neutrinos whose wavepackets travel with same energy $E$ and whose momenta may be Taylor expanded as $p_i=E-m_i^2/2E$, the probability of oscillation to flavor $\nu_\beta$ from flavor $\nu_\alpha$ is
\beq  \label{genform}
P_{\nu_\alpha \to \nu_\beta}=\sum_{i,j} U_{\alpha i}^* U_{\beta i}  U_{\alpha j}  U_{\beta j}^* \exp \left[ i(m_i^2-m_j^2) L/2E \right]
\eeq
This formula is easy to evaluate in the limit of many light mass eigenstates because unitarity simplifies the evaluation of the sum. However, the fifth neutrino may either not be accessible or may be produced in a reduced phase space. In this case the evaluation of the sum is less straightforward because the assumption that production processes for all mass eigenstates are similar may no longer be good. With the definitions ${\mathbb U}_{\alpha \beta i j} \equiv U_{\alpha i}^* U_{\beta i} U_{\alpha j} U_{\beta j}^*$ and $x_{ij} \equiv (m_i^2-m_j^2) L/4E$ and taking a phase space suppression factor $0<a<1$ on oscillations through $n_5$, the oscillation probability is
\beq \begin{array}{rcl} \label{first}
P_{\nu_\alpha \to \nu_\beta}
&=&\sum_{i,j}{\mathbb U}_{\alpha \beta i j} \exp( 2 i x_{ij} ) - 2 \Re \left[ (1- a) \sum_{j}{\mathbb U}_{\alpha \beta 5 j}  \exp(2  i x_{ij} ) \right]+ (1- a) {\mathbb U}_{\alpha \beta 5 5}  .
\end{array} \eeq
Carrying out some standard simplifications allows us to write Eq.~\eqref{first} as
\beq \begin{array}{rcl}
P_{\nu_\alpha \to \nu_\beta} &=& \delta_{\alpha \beta} \left[ 1  -  2(1 -a) |U_{\alpha 5}| |U_{ \beta 5}| \right] + (1-a) |U_{\alpha 5}|^2 |U_{\beta 5}|^2 - 4  \sum_{5>i>j} \Re[{\mathbb U}_{\alpha \beta i j}] \sin^2 x_{ij}    \\ && ~~~ - 4 a  \sum_{j=1}^4 \Re[{\mathbb U}_{\alpha \beta 5 j}]  \sin^2 x_{5j}  - 2 \sum_{5>i>j} \Im[{\mathbb U}_{\alpha \beta i j}] \sin 2 x_{ij}  - 2 a \sum_{j=1}^4 \Im[{\mathbb U}_{\alpha \beta 5 j}]  \sin 2 x_{5j}\label{oscform}
\end{array} \eeq
where we use the unitarity condition $\sum_i U_{\alpha i} U^*_{\beta i} = \delta_{\alpha \beta}$. In the limit $ a \to 1$ we recover  the standard result.

In all of the experiments of interest we may ignore oscillations due to the mass differences $\Delta m_{12}^2$ and $\Delta m_{23}^2$, and Eq.~\eqref{oscform} simplifies. The oscillation probability of interest in appearance searches such as LSND is found to be
\beq \label{genapp}
P_{\nu_e \to \nu_\mu}  = |U_{e 4}|^2 |U_{\mu 4}|^2 \left\{ a [(1-r)^2 +4 r \sin^2 \beta ]  + (1-r)^2 +4r \sin^2 (x_{41} \pm \beta)  \right\},
 \eeq
where + $(-)$ is for $\nu$ ($\bar{\nu}$) oscillations, and the definitions of $r$ and $\beta$ are given in Eq.~\eqref{rdef} in the text.  For disappearance experiments, the relevant formula is
\beq \begin{array}{rcl}
1-P_{\nu_\alpha \to \nu_\alpha} &=&  \sin^2 2\theta_{\alpha 4} \sin^2 x_{41} + 2 |U_{\alpha 5} |^2(1- \frac{a+1}{2}|U_{\alpha 5} |^2)  \label{gendisapp},
\end{array} \eeq
where $\sin^2 2\theta_{\alpha 4}=4|U_{\alpha 4} |^2(1-|U_{\alpha 4} |^2 -|U_{\alpha 5} |^2)$ and we assume that oscillations through $n_5$ are averaged. We emphasize that because the phase space factor $a$ only enters at second order in $|U_{\alpha 5} |^2$ the phase space available to $n_5$ has very little impact on the predictions and constraints of disappearance experiments.

The ``zero-distance" effect \cite{Langacker:1988ur} arises because even in the limits $a \to 0$ and $x_{41} \to 0$ there remains a nonzero probability for oscillation. Another consequence of the very heavy state is that for fixed $\alpha$ the sum of  Eq.~\eqref{oscform} over $\beta$ betrays nonunitarity. This indicates that we have normalized our states incorrectly. However, this is cancelled by an inverse change in the production and detection cross-sections in the types of experiments considered here \cite{Antusch:2006vwa}, so we may use the formulae as if the probabilities were unitary.

\end{document}